\documentclass[aps,prd,twoside,twocolumn,nofootinbib,10pt,showpacs,floatfix]{revtex4-1}
\usepackage{amsmath,amssymb}
\usepackage{graphicx,bm}
\usepackage{slashed}
\usepackage{epstopdf}
\usepackage{ulem} %% for strike-through
\usepackage[usenames]{color}
\usepackage{float}
\usepackage{hyperref}
\usepackage{subfigure}
\usepackage{subfigure}
\usepackage{rotating}
\usepackage{color}
\usepackage{multirow}
\usepackage{dcolumn}
\usepackage{overpic}
\usepackage{booktabs}
\usepackage{makecell}
\usepackage{diagbox}

\renewcommand\sout{\bgroup  \ULdepth=-.5ex \ULset}

\newsavebox{\tablebox}
\begin{document}
%\begin{CJK}{GBK}{}

\title{Improved understanding of the peaking phenomenon existing in the new di-$J/\psi$ invariant mass spectrum from the CMS Collaboration}
\author{Jun-Zhang Wang$^{1,2}$}\email{wangjzh2012@lzu.edu.cn}
\author{Xiang Liu$^{2,3,4,5}$\footnote{Corresponding author}}\email{xiangliu@lzu.edu.cn}
\affiliation{$^1$School of Physics and Center of High Energy Physics, Peking University, Beijing 100871, China\\
$^2$School of Physical Science and Technology, Lanzhou University, Lanzhou 730000, China\\
$^3$Research Center for Hadron and CSR Physics, Lanzhou University $\&$ Institute of Modern Physics of CAS, Lanzhou 730000, China\\
$^4$Lanzhou Center for Theoretical Physics, Lanzhou University, Lanzhou 730000, China\\
$^5$Key Laboratory of Theoretical Physics of Gansu Province, and Frontiers Science Center for Rare Isotopes, Lanzhou University, Lanzhou 730000, China}

\date{\today}

\begin{abstract}

Very recently, the CMS Collaboration reported a peaking phenomenon existing in the di-$J/\psi$ invariant mass spectrum from $pp$ collision, by which the $X(6900)$ structure announced by the LHCb Collaboration was confirmed, but also more enhancement structures were discovered. Facing such a novel phenomenon, 
in this work we indicate that these new features reflected from the CMS measurement provide a good implication for a dynamical mechanism which reproduces the novel peaking phenomenon in the reported $J/\psi$-pair mass spectrum well. This mechanism is due to special reactions, where different charmonium pairs directly produced by $pp$ collision may transit into the final state of $J/\psi J/\psi$. The present work provides a special viewpoint to decode these observed fully charm structures in the $J/\psi$-pair invariant mass spectrum. 
\end{abstract}

\maketitle

\section{Introduction}\label{sec1}

%{\noindent\it Introduction.---}

The discovery of a series of new hadron states has stimulated theorists' extensive interest in the study of exotic hadronic configuration with more constituent quarks and gluons since 2003. With continuous efforts from experimentalists and theorists, more and more manifestly exotic structures have been identified, which include hidden-charm pentaquark  $P_c(4312)^+$, $P_c(4440)^+$ , $P_c(4457)^+$  \cite{LHCb:2019kea}, a series of charmoniumlike $XYZ$ states as hidden-charm tetraquark candidates \cite{Belle:2003nnu,BESIII:2013ris,BESIII:2013ouc,LHCb:2021uow}, doubly charmed tetraquark $T_{cc}^+$ \cite{LHCb:2021auc},  and so on (see review articles for more progress \cite{Chen:2016qju,Liu:2019zoy,Chen:2022asf,Guo:2017jvc,Olsen:2017bmm,Brambilla:2019esw} ). Therefore, how to decode the nature of these exotic hadronic matters has become an extremely interesting research topic in hadron physics.

Different from other new hadronic states, the $X(6900)$ was first reported in the di-$J/\psi$ invariant mass spectrum from LHCb \cite{LHCb:2020bwg}, where its minimal quark constituent is fully charm $cc\bar{c}\bar{c}$. 
At the ICHEP 2022 conference, the ATLAS Collaboration reported the evidence of a four-charm-quark excess \cite{Atlas2022}. Because of the absence of light flavor degrees of freedom, the $X(6900)$ structure can provide a unique platform to study the dynamics inside mulitibody heavy flavor system.  In addition to the discovery of the di-$J/\psi$ structure around 6.9 GeV,  LHCb's data also imply the existence of a broad enhancement structure near the production threshold of the $J/\psi$ pair \cite{LHCb:2020bwg}. Thus, these novel fully heavy enhancement  phenomena have inspired numerous theoretical explanations, which include the mainstreaming fully charm tetraquark states (compact or diquark type) \cite{Chen:2020xwe,Jin:2020jfc,Wang:2020ols,Lu:2020cns,Yang:2020rih,Deng:2020iqw,Chen:2020lgj,Albuquerque:2020hio,Sonnenschein:2020nwn,Giron:2020wpx,Richard:2020hdw,Becchi:2020uvq,liu:2020eha,Bedolla:2019zwg,Chao:2020dml,Karliner:2020dta,Faustov:2020qfm,Gordillo:2020sgc,Weng:2020jao,Zhang:2020xtb,Yang:2020wkh,Zhao:2020zjh,Faustov:2021hjs,Ke:2021iyh,Yang:2021hrb,Li:2021ygk,Asadi:2021ids,Kuang:2022vdy,Wu:2022qwd,Wang:2021kfv}, the dynamical mechanism from the  scattering of double charmonia \cite{Wang:2020wrp,Dong:2020nwy,Guo:2020pvt,Liang:2021fzr,Dong:2021lkh,Zhuang:2021pci}, the gluonic tetracharm hybrid \cite{Wan:2020fsk}, a Higgs-like boson \cite{Zhu:2020snb}, {\it etc}.   Additionally,  the production property and decay behavior of fully charm tetraquark states were also discussed by several research groups \cite{Maciula:2020wri,Ma:2020kwb,Feng:2020riv,Zhu:2020xni,Feng:2020qee,Gong:2020bmg,Goncalves:2021ytq}. We should pay more experimental and theoretical efforts to clarify the nature of peaking phenomenon in the di-$J/\psi$ mass distribution.

Very recently, the CMS Collaboration released their measurements on the $J/\psi$-pair mass spectrum by using $135$~fb$^{-1}$ proton-proton data at center of mass energies of 13 TeV \cite{Cms2022},  which not only confirmed the existence of the $X(6900)$ reported by LHCb with significance 9.4$\sigma$, but also found signals of some new peaking structures. By using a relativistic $S$-wave Breit-Wigner function, the resonance parameters of two new structures, the $X(6600)$ and $X(7300)$, were obtained \cite{Cms2022}
\begin{eqnarray}
m_{X(6600)}&=&6552\pm10\pm12 ~ \mathrm{MeV},~\nonumber \\ \Gamma_{X(6600)}&=&124 \pm 19\pm 34 ~ \mathrm{MeV}, \nonumber \\
m_{X(7300)}&=&7287\pm19\pm5 ~ \mathrm{MeV},~ \nonumber \\ \Gamma_{X(7300)}&=&95 \pm 46\pm 20 ~ \mathrm{MeV}. \nonumber
\end{eqnarray}
%where a narrow structure around 6.9 GeV was observed with a significant signal of more than standard deviation of 5.1$\sigma$ \cite{Aaij:2020fnh}. Besides, there exist an obvious broad structure ranging from the threshold of di-$J/\psi$ to 6.8 GeV, and an underlying peak near 7.3 GeV. 
Undoubtedly, the new CMS measurement can provide more refined hints to decode the novel peaking phenomena appearing in the double $J/\psi$ mass spectrum.

Inspired by the new experimental results from CMS, in this work, we further apply a dynamical mechanism to understand these observed fully charm enhancement structures. Here, the adopted 
dynamical mechanism 
is based on a special reaction that different charmonium pairs from direct hadroproduction may transit into final states of di-$J/\psi$ \cite{Wang:2020wrp}. This dynamical mechanism has succeed in explaining the $X(6900)$ as a threshold cusp structure resulting from the intermediate $\chi_{c0}\chi_{c1}$ scattering \cite{Wang:2020wrp}.  As suggested in Ref. \cite{Guo:2014iya}, we first extend this dynamical mechanism by considering higher-order multiloop contributions.
By applying the extended dynamical model to fit the line shape of the di-$J/\psi$ mass distribution measured by both CMS and LHCb, we demonstrate that in addition to the intermediate $J/\psi \psi(3686)$ and $\chi_{c0}\chi_{c1}$ scattering which are enough to explain the LHCb data,  the contributions of the remaining two allowed characteristic intermediate channels $\chi_{c1}\eta_c$ and $\chi_{c2}\chi_{c2}$ in the measured energy region can be explicitly revealed by the features appearing in the CMS measurement, which can correspond to the newly observed fully charm enhancement structure around 6.6 and 7.3 GeV, respectively. Thus, this is direct evidence to support this dynamical production mechanism based on a double charmonia scattering. Furthermore, we discuss the origins of these di-$J/\psi$ peaking structures in the dynamical mechanism under two fitting schemes. The present study is helpful to improve the understanding of the peaking phenomena in the di-$J/\psi$  invariant mass spectrum.

This paper is organized as follows. After the Introduction, in Sec. \ref{sec2}, we present the theoretical framework of the extended dynamical mechanism of producing the fully charm enhancement structures in the di-$J/\psi$ mass spectrum. In Sec. \ref{sec3}, we discuss the improved understanding for the peaking phenomena in the double $J/\psi$ mass spectrum based on the CMS measurement. Finally, this paper ends with the discussion and conclusion in Sec. \ref{sec4}.

\section{The extended dynamical mechanism}\label{sec2}

%{\noindent\it The extended dynamical mechanism}.---

The hadroproduction of  double charmonium in a high-energy proton-proton collider is usually achieved by the $gg\to (c\bar{c})(c\bar{c}) + X$ process in single parton scattering (SPS) \cite{Sun:2014gca,Likhoded:2016zmk,Baranov:2011zz,Lansberg:2013qka,Lansberg:2014swa,Lansberg:2015lva,Shao:2012iz,Shao:2015vga}  and the $gggg\to (gg\to c\bar{c}) (gg \to c\bar{c}) + X$ in double parton scattering (DPS) \cite{Calucci:1997ii,Calucci:1999yz,DelFabbro:2000ds,Lansberg:2019adr}. This production mechanism usually plays a main role for producing the continuum distribution in the invariant mass spectrum of double charmonium, whereas the observed novel enhancement phenomenon in the di-$J/\psi$ mass spectrum by recent LHCb and CMS measurement tells us that  a new origin of double $J/\psi$ hadroproduction should exist, which has stimulated some theoretical discussions on a dynamical production mechanism of the $J/\psi$ pair \cite{Wang:2020wrp,Dong:2020nwy,Guo:2020pvt,Liang:2021fzr,Dong:2021lkh,Zhuang:2021pci,Wang:2020tpt}.

\begin{figure}[b]
	\includegraphics[width=8.6cm,keepaspectratio]{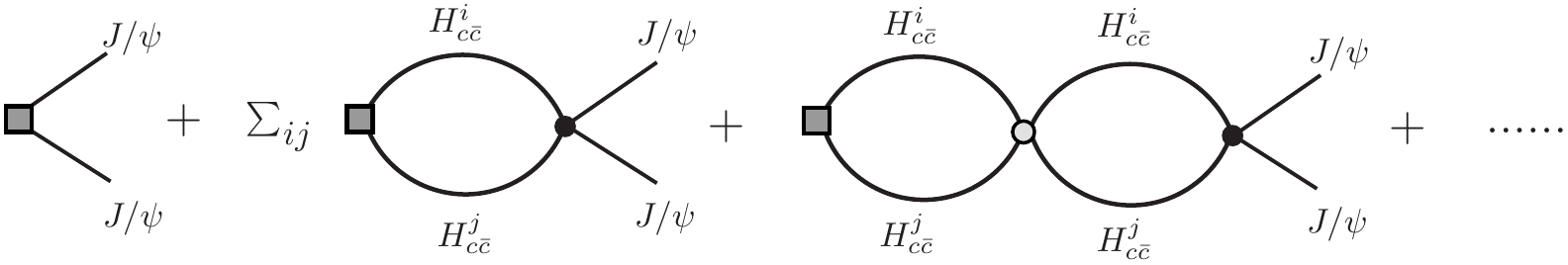}
	\caption{ The schematic diagrams for the dynamical production mechanism of a double $J/\psi$, where $H^{i}_{c\bar{c}}$ stands for allowed intermediate charmonium states. %such as $\eta_c$, $J/\psi$, $\chi_{cJ}$ with $J=0,1,2$, etc. 
Here, the gray rectangle represents the direct production of a double charmonium in hadron colliders. \label{fig:feynman}  }
\end{figure}

This new dynamical mechanism is based on a special reaction, where various combinations of double charmonia that are directly produced via both the  SPS and DPS processes are transferred into final states of $J/\psi J/\psi$. The combination selection of an intermediate charmonium pair depends on the quantum number conservation of  the reaction system. This dynamical mechanism was proposed in our previous work for the first time and has been found to produce the $X(6900)$ structure well \cite{Wang:2020wrp}. In this work, we further extend this reaction picture to the multiloop case, where the higher-order coupling contribution of an intermediate double charmonia scattering to the same intermediate double charmonia will be taken into account. These contributions have been demonstrated to provide more abundant dynamical information for the intermediate scattering process \cite{Guo:2014iya}.  The corresponding schematic diagrams are presented in Fig. \ref{fig:feynman}, where the interaction between an intermediate charmonium pair and a transferred  double charmonium is absorbed into a vertex. Here, it is worth mentioning that we have to ignore the coupled channel effects in the subsequent analysis  because of the absence of  relevant experimental data.

Concrete theoretical calculations on the direct production of  double charmonium in high-energy proton-proton collisions by the SPS and DPS mechanisms  are quite challenging \cite{Sun:2014gca,Likhoded:2016zmk,Baranov:2011zz,Lansberg:2013qka,Lansberg:2014swa,Lansberg:2015lva,He:2019qqr,He:2015qya,Lansberg:2020rft,Lansberg:2019fgm,Li:2009ug}. Fortunately,  here we focus only  on the line shape of their contribution  in the di-$J/\psi$ invariant mass spectrum. So the $S$-wave direct production amplitude of a double charmonium $H_{c\bar{c}}^i H_{c\bar{c}}^j$ marked in Fig. \ref{fig:feynman} can be parametrized as \cite{Wang:2020wrp}
\begin{eqnarray}
\mathcal{A}_{direct}^{ij}=\left[(g^{ij}_{direct})^2~e^{c_0m_{ij}}~\frac{1}{8\pi}\frac{\sqrt{\lambda(m_{ij}^2,m_{i}^2,m_{j}^2)}}{m_{ij}^2}\right]^{1/2},~ \label{eq:1}
\end{eqnarray}
where $m_{ij}$ is the invariant mass of  double charmonia $H_{c\bar{c}}^i H_{c\bar{c}}^j$. The K$\ddot{\textrm{a}}$llen function is defined to be $\lambda(x,y,z)=x^2+y^2+z^2-2xy-2xz-2yz$. For the one-loop process with an $S$-wave interaction between intermediate charmonium pairs shown in  Fig. \ref{fig:feynman}, the scattering amplitude of producing a $J/\psi$$J/\psi$  becomes the one proportional to the scalar two-point loop integral, whose analytical form within a nonrelativistic form can be given by, in the rest frame of di-$J/\psi $,
\begin{eqnarray}
&&L_{ij}(m_{J/\psi J/\psi})\nonumber\\&&
\!=\!\int \frac{dq^4}{(2\pi)^4} \frac{i2\sqrt{2}e^{-(2\vec{q}~)^{2}/\alpha^2}}{(q^2-m_{i}^2+i\epsilon)((P-q)^2-m_{j}^2+i\epsilon)} \nonumber \\
&&\!\simeq\!\frac{-1}{4m_{i}m_{j}}
\!\left\{\!\frac{-2\mu\alpha}{(2\pi)^{3/2}}\!+\!\frac{2\mu\sqrt{2\mu m_0}\left(\textrm{erfi}\left[\frac{\sqrt{8\mu m_0}}{\alpha}\right]\!-\!i\right)}{2\pi/e^{-\frac{8\mu m_0}{\alpha^2}}} 
\!\right\}\!, \label{eq:2}\nonumber\\
\end{eqnarray}
where $\mu=(m_im_j)/(m_i+m_j)$, $m_0=m_{J/\psi J/\psi}-m_{i}-m_{j}$, and $m_i$($m_j$) is the hadron mass of an intermediate charmonium  $H_{c\bar{c}}^i$($H_{c\bar{c}}^j$). Here, an exponential form factor $e^{-(2\vec{q}~)^{2}/\alpha^2}$ with a cutoff parameter $\alpha$ is introduced to avoid the ultraviolet divergence of scalar two-point loop integral.  In our previous work \cite{Wang:2020wrp}, the intermediate charmonium pairs  $H_{c\bar{c}}^iH_{c\bar{c}}^j$= $J/\psi J/\psi$, $\eta_c\chi_{cJ}$, $J/\psi h_c$ and $\chi_{cJ}\chi_{cJ}$ with $J=0,1,2$ were considered, which are the charmonium combination composed of radially ground states and completely cover the concerned energy region  from 6.194 to 7.300 GeV. Here,  important evidence supporting these combinations is that  the direct hadroproduction rates of these radially ground charmonia have been proved to be comparable with that of $J/\psi$ by both experiments \cite{Aaij:2014bga,Aaij:2011sn} and theoretical estimations from nonrelativistic QCD \cite{Bodwin:1994jh,Ma:2014mri,Li:2011yc,Butenschoen:2014dra,Han:2014jya,Bodwin:2015iua,Ma:2010vd,Artoisenet:2009wk,Butenschoen:2013pxa}. Of course, if one further extends the selection criterion to radially exciting charmonium states, there should be two more channels $J/\psi \psi(3686)$ and $J/\psi\psi(3770)$ in the same energy region, whose possible contributions will be also investigated in this work. It is worth noticing that we include the width effects of intermediate charmonium states by replacing $m_{i}$ in Eq. (\ref{eq:2}) with ($m_{i}-i\Gamma_{i}/2$).

When considering all combinations of  a charmonium pair, one can find that there exist 13 intermediate channels in the di-$J/\psi$ energy region from 6.194 to 7.300 GeV. Obviously, it is impossible to include so many dynamical reactions in a practical theoretical analysis.  Fortunately, we found that their threshold positions are mainly concentrated in five local energy regions,  which provides us convenience to deal with this problem. Considering the fact that the contributions of intermediate channels in the same local energy region may overlap and then behave like one peak structure, we can choose a representative channel to absorb the contributions from other nearby scattering channels. Thus, all possible dynamical scattering channels for double $J/\psi$ hadroproduction in the energy range below 7.3 GeV include $J/\psi J/\psi$, $\eta_c\chi_{c1}$, $J/\psi \psi(3686)$, $\chi_{c0}\chi_{c1}$, and $\chi_{c2} \chi_{c2}$.

The two-loop and three-loop production amplitudes in Fig. \ref{fig:feynman} are proportional to $\mathcal{C}_{ij}L_{ij}(m_{J/\psi J/\psi})^2$ and $-\mathcal{C}_{ij}^2L_{ij}(m_{J/\psi J/\psi})^3$, respectively, and higher-order loop contributions can be written accordingly, where $\mathcal{C}_{ij}$ represents the coupling strength of intermediate charmonium pairs scattering to the same double charmonia. If we sum up all loop diagram contributions, the line shapes of the invariant mass distribution of di-$J/\psi$ caused by the extended dynamical mechanism are given by
\begin{eqnarray}
\mathcal{A}^2_{ij}(m_{J/\psi J/\psi})\!=\!\frac{g_{ij}^2 L_{ij}(m_{J/\psi J/\psi})^2}{(1\!+\!\mathcal{C}_{ij}L_{ij}(m_{J/\psi J/\psi}))^2}\frac{e^{c_0m_{J/\psi J/\psi}}p_{J/\psi}}{m_{J/\psi J/\psi}}~~~ \label{eq:3}
\end{eqnarray}
and
\begin{eqnarray}
\mathcal{A}^{\prime2}_{ij}(m_{J/\psi J/\psi})\!=\! \frac{g_{ij}^2 L_{ij}(m_{J/\psi J/\psi})^2}{(1\!+\!\mathcal{C}_{ij}L_{ij}(m_{J/\psi J/\psi}))^2}\frac{e^{c_0^{\prime}m_{J/\psi J/\psi}}p_{J/\psi}^3}{m_{J/\psi J/\psi}}~~~ \label{eq:4}
\end{eqnarray}
for two types of system parity  $P=+$ and $P=-$, respectively, in which $p_{J/\psi}$ is the momentum of a final state $J/\psi$, and the factor $e^{c_0^{(\prime)}m_{J/\psi J/\psi}}=e^{c_0^{(\prime)}m_{ij}}$ is introduced to describe the energy dependence of the direct hadroproduction amplitude of intermediate charmonium pairs $H_{c\bar{c}}^i H_{c\bar{c}}^j$ as shown in Eq. (\Ref{eq:1}). {Here, it is worth noting that the coupling constant $g_{ij}$ involves two contributions, the production ratio and
the transition from $H_{c\bar{c}}^i H_{c\bar{c}}^j$ to the di-$J/\psi$ channel.} The $c_0=-1.5$ and $c_0^{\prime}=-1.0$  are extracted by fitting the LHCb data \cite{Wang:2020wrp,LHCb:2020bwg}. Here, the involved intermediate channels $H_{c\bar{c}}^i H_{c\bar{c}}^j=J/\psi J/\psi, \chi_{cJ}\chi_{cJ},  J/\psi \psi(3686), J/\psi\psi(3770)$ and $H_{c\bar{c}}^i H_{c\bar{c}}^j=\eta_c\chi_{cJ}, J/\psi h_c$ are related to the parity-even and parity-odd, respectively, and the parity-odd amplitude  corresponds to the hadroproduction of a $P$-wave double $J/\psi$.

\begin{figure}[t]
\includegraphics[width=8.0cm,keepaspectratio]{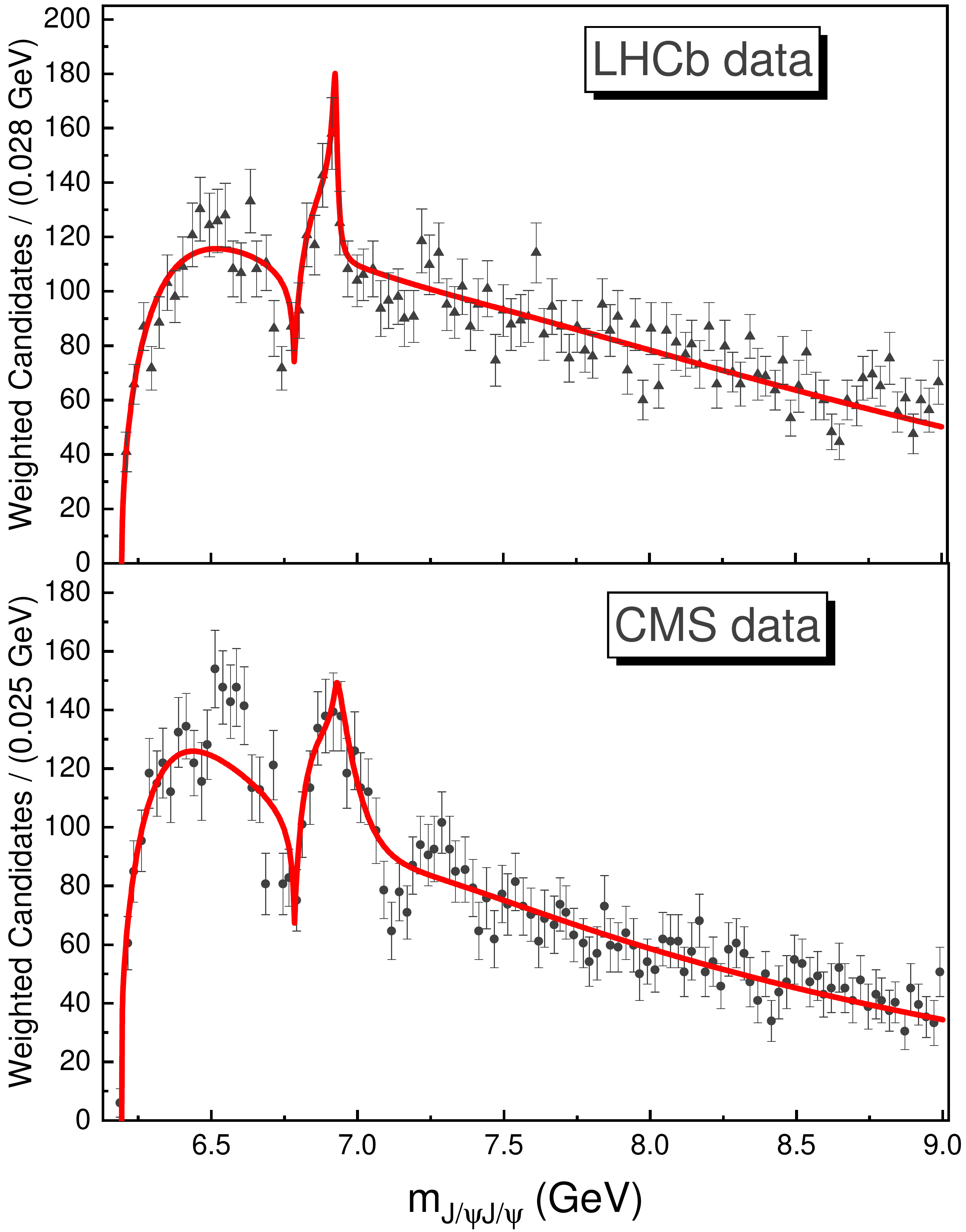}
	\caption{ The comparison between the fitting results to the LHCb and new CMS data based on an extended dynamical mechanism, where two parity-even intermediate rescattering  channels  $J/\psi \psi(3686)$ and $\chi_{c0}\chi_{c1}$ are included.   \label{fig:Fitexp-minimumpara}  }
\end{figure}

%Without considering the coupled channel effects among different intermediate double charmonium channels, 

The line shape for the total invariant mass spectrum of producing a double charmonium $J/\psi J/\psi$ in high-energy proton-proton colliders can be written as
\begin{eqnarray}
&\mathcal{A}^2\!=\mid\! \mathcal{A}_{direct}^{J/\psi J/\psi}(m_{J/\psi J/\psi})\!+\!\sum_{mn}e^{i\phi^{mn}}\mathcal{A}_{mn}(m_{J/\psi J/\psi})\! \mid^2  \nonumber \\
&\quad+\!\!\mid\! \mathcal{A}_{direct}^{J/\psi J/\psi \prime}(m_{J/\psi J/\psi})\!+\!\sum_{mn}e^{i\phi^{mn}}\mathcal{A}^{\prime}_{mn}(m_{J/\psi J/\psi})\! \mid^2, \nonumber \\
\end{eqnarray}
where $\phi^{mn}$ is the phase between direct contribution and the corresponding rescattering dynamical  process, and a background term $\mathcal{A}_{direct}^{J/\psi J/\psi \prime}=(\frac{g^{\prime2}_{direct}\lambda(m_{J/\psi J/\psi}^2,m_{J/\psi}^2,m_{J/\psi}^2)^{\frac{1}{2}}e^{c_0^{\prime}m_{J/\psi J/\psi}}p_{J/\psi}^2}{8\pi m_{J/\psi J/\psi}^2})^{\frac{1}{2}}$ describes the direct production of double $J/\psi$ with $P=-1$. { Generally, the coupled channel effect in $T$ matrix may bring some imaginary part term for amplitude, which can be partially absorbed in  phase angle factor $e^{i\phi^{mn}}$.  In addition,  there may exist an inherent phase difference between the production vertex of different intermediate charmonium pairs. Here, it is worth mentioning that the $J/\psi J/\psi$ spectrum production in high-energy $pp$ collision is from a complex inclusive reaction $pp \to J\psi J/\psi X$, where the unitary may be violated.  } Here, we choose only the phase angle $\phi^{mn}=0$ and $\pi$ for the consideration of reducing the fitting parameters, which can relate to the constructive and destructive situation, respectively.

In the extended dynamical mechanism, there exists an integral singularity at the threshold of ($m_{i}+m_{j}$) appearing at the on shell of two intermediate charmonia because of a square root branch point $\sqrt{m_{J/\psi J/\psi}-m_{i}-m_{j}}$ from the scattering amplitude. From Eqs. (\ref{eq:3}) and (\ref{eq:4}), it can be seen that the extended dynamical amplitude returns to the single loop situation when $\mathcal{C}_{ij}L_{ij}<<1$ as shown in Ref. \cite{Wang:2020wrp}, in which this integral singularity of amplitude will cause a nonresonant cusp line shape on the distribution of the $m_{J/\psi J/\psi}$ and its peak position is almost exactly at the threshold of inducing the on-shell intermediate charmonium pairs.  Interestingly, if the coupling constant $\mathcal{C}_{ij}$ is strong enough, the threshold cusp may transform into a pole structure, %which can correspond to a fully-charm tetraquark state from the interaction between intermediate charmonium pairs, 
whose configuration should be quite novel and is totally different from the compact type of $cc\bar{c}\bar{c}$.

\section{What can we learn from the  di-$J/\psi$ spectrum reported by  CMS?}\label{sec3}

%{\it What can we learn from the CMS's di-$J/\psi$ spectrum?}---
With the above preparations, we can study  the di-$J/\psi$ mass spectrum based on the extended dynamical model.  In our previous research work, we have indicated that an underlying broad enhancement near threshold and the reported $X(6900)$ structure in LHCb data can be reproduced well by the parity-odd channel $\chi_{c1}\eta_c$ and  parity-even channel  $\chi_{c0}\chi_{c1}$, respectively \cite{Wang:2020wrp}. However, the authors in Ref. \cite{Dong:2020nwy} found that the broad enhancement near threshold can also be explained by the destructive contribution from the parity-even channel $J/\psi\psi(3686)$, whose threshold just exactly locates at an obvious dip position between two peaking structures.  This fact means that the LHCb data cannot distinguish these two channel contributions. Fortunately,  the new CMS measurement can reveal more critical information to clarify this problem, where more details on the line shape of novel fully charm enhancement structures were presented compared with previous LHCb data.

By checking the relevant experimental data of CMS, we found a new data accumulation in the vicinity of 6.6 GeV, which is not far from the threshold of the $\chi_{c1} \eta_c$ channel. In order to demonstrate that this accumulation may be caused by the parity-odd contribution of $\chi_{c1} \eta_c$ instead of  $J/\psi \psi(3686)$, we performed an independent fit to the CMS and LHCb data by considering two parity-even intermediate channels  $J/\psi \psi(3686)$ and $\chi_{c0}\chi_{c1}$, whose fitting results are shown in Fig.~\ref{fig:Fitexp-minimumpara}.  It can be seen that the line shape of  the di-$J/\psi$ invariant mass spectrum measured by LHCb can be reproduced well in the present scenario, which is consistent with the conclusion of Ref. \cite{Dong:2020nwy}. However, it is apparent that the CMS data cannot be described well in the same scheme, especially for two energy regions with a large divergence. The first one is the energy range between 6.6 and 6.7 GeV, which just corresponds to the threshold of the rescattering  channel  $\chi_{c1} \eta_c$. The second region is around 7.1 GeV. As a matter of fact, our former work has given the predictions for the existence of possible fully charm structures in this energy region although the LHCb experiment does not show any obvious hints  \cite{Wang:2020wrp,LHCb:2020bwg}, where  $\chi_{c1}\chi_{c1}$, $\chi_{c1}\chi_{c2}$ and $\chi_{c2}\chi_{c2}$ contribute to the energy position of ($7.03-7.13$) GeV. Hence, in the following, we will explore whether the CMS data  on this critical energy region  of the double $J/\psi$ mass spectrum can be reproduced by the inclusion of two new intermediate double charmonium channels $\chi_{c1} \eta_c$ and $\chi_{c2}\chi_{c2}$.

\begin{figure}[t]
\includegraphics[width=8.7cm,keepaspectratio]{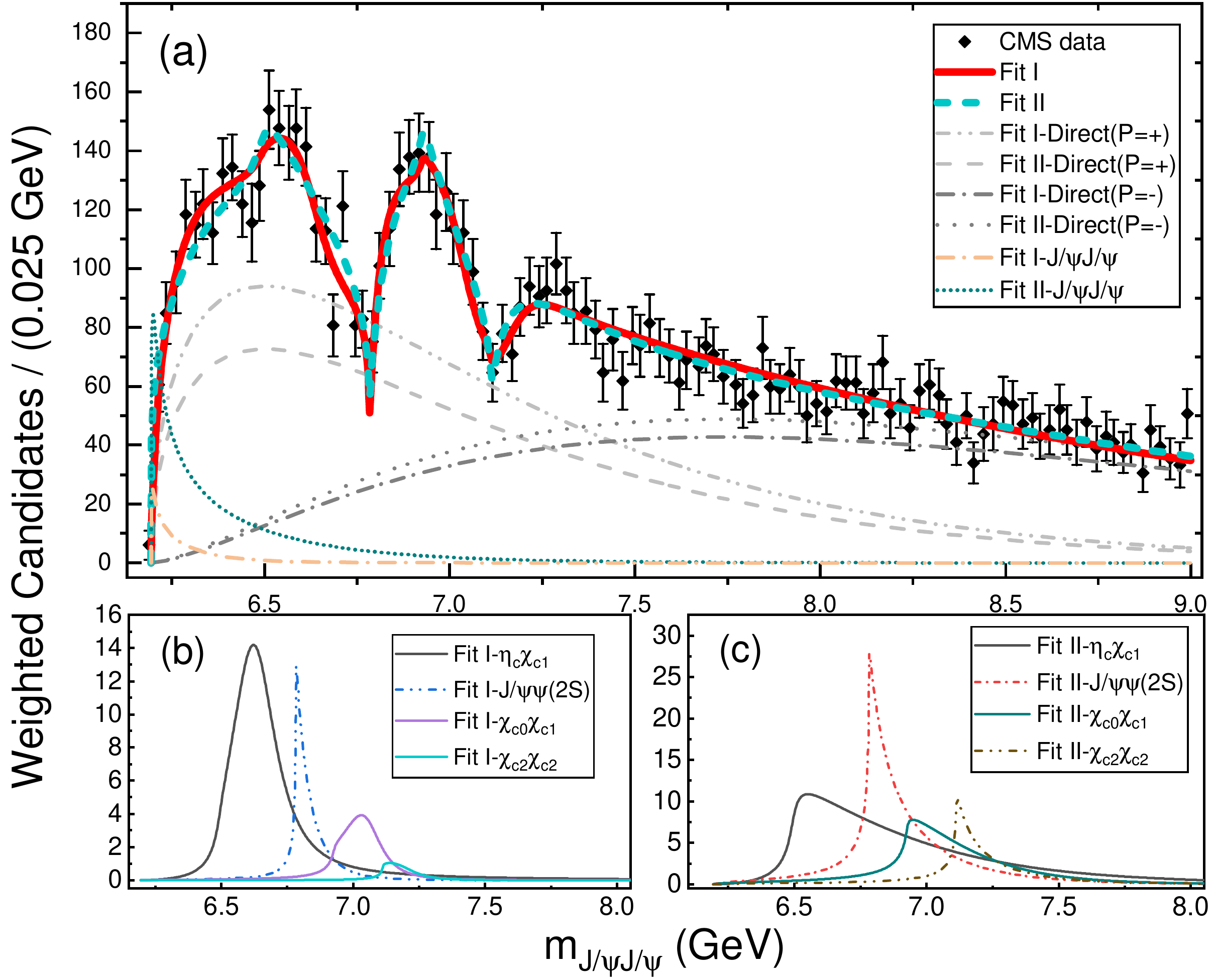}
	\caption{ The complete theoretical fit to the invariant mass distribution of $J/\psi$ pair measured by the CMS Collaboration \cite{Cms2022} within an extended dynamical mechanism. Here, two fitting schemes of fit I and fit II are presented, in which four characteristic intermediate channels  $\eta_c\chi_{c1}$, $J/\psi\psi(3686)$, $\chi_{c0}\chi_{c1}$ and $\chi_{c2}\chi_{c2}$ are introduced.  \label{fig:Fitexp}  }
\end{figure}

Based on the extended dynamical production mechanism, the complete theoretical analysis of the CMS experimental data of the invariant mass spectrum of $J/\psi J/\psi $ is presented in  Fig. \ref{fig:Fitexp}. And the corresponding fitted parameters are summarized in Table \ref{table:parameter}. One can see that the novel peaking structures shown in the CMS data can be reproduced well by the red solid line in the fit-I scheme, where the fitting $\chi^2/d.o.f.=0.657$ is obtained. Specifically, the experimental data around 6.6 and 7.1 GeV can indeed be described by the contribution from the $\chi_{c1} \eta_c$ and $\chi_{c2}\chi_{c2}$ channel, respectively.  This means that the CMS measurement provides definite evidence for confirming the contribution of parity-odd channel $\chi_{cJ} \eta_c$ in the $J/\psi$-pair mass spectrum for the first time. From Table \ref{table:parameter}, one can find that the central values of the fitted coupling constants $\mathcal{C}_{\chi_{c1} \eta_c}=342$,  $\mathcal{C}_{\chi_{c0} \chi_{c1}}=380$ and $\mathcal{C}_{\chi_{c2} \chi_{c2}}=145$ with  cutoff $\alpha=0.871$ are relatively larger than $\mathcal{C}_{J/\psi \psi(3686)}=-20$ in the fit-I scheme. A such a strong interaction may convert the threshold singularity to a dynamically generated pole structure.  By performing a corresponding pole analysis, we found that the $J/\psi \psi(3686)$ channel produces a threshold cusp structure and three resonance poles from the rescattering channels $\chi_{c1} \eta_c$, $\chi_{c0} \chi_{c1}$ and $\chi_{c2} \chi_{c2}$ really appear in the second Riemann sheet, whose pole positions are determined to be
\begin{eqnarray}
\mathcal{E}_{\chi_{c1} \eta_c}=(6.625-0.107 i) ~~\mathrm{GeV}, \nonumber \\
\mathcal{E}_{\chi_{c0} \chi_{c1}}=(7.050-0.089 i) ~~\mathrm{GeV}, \\
\mathcal{E}_{\chi_{c2} \chi_{c2}}=(7.170-0.108 i) ~~\mathrm{GeV},\nonumber 
\end{eqnarray}
respectively. Their individual line shape on the invariant mass spectrum of double $J/\psi$ can be found in Fig. \ref{fig:Fitexp} (b), in which there is an obvious line shape difference between cusp and resonance solution.
Although the resonance solution is from the best $\chi^2$ fitting for the experimental data, we must point out that it is difficult to understand such strong coupling constant $\mathcal{C}$ for a double charmonium scattering to charmonium pairs, where the long-distance interaction from the direct exchange of light medium mesons should be relatively suppressed due to the absence of light quark freedom in the fully heavy reaction system. Here,  some interesting unknown nonperturbative dynamics should play important role, which may be reliably revealed by lattice QCD in the future.

\begin{table}[t]
  	\caption{The fitted parameters for reproducing the line shape of the CMS data within the fit-I and fit-II schemes.}
  	\setlength{\tabcolsep}{2.6mm}{
  	\begin{tabular}{lccccccc}
			\toprule[1.0pt]
     Parameters &  Fit I  &  Fit II  \\
			\toprule[1.0pt]
          $\mid g_{direct}^{\prime}/g_{direct} \mid$  & $0.0575\pm0.0009$  & $0.0699\pm0.0010$    \\ 
            $ g_{\eta_c\chi_{c1}}/g_{direct}^{\prime} $  & $125\pm26$  & $28.2\pm2.7$    \\
            $ g_{J/\psi\psi(3686)}/g_{direct} $  & $-26.1\pm4.7$  & $-16.4\pm1.7$    \\
             $ g_{\chi_{c0}\chi_{c1}}/g_{direct} $  & $32.9\pm5.8$  & $16.2\pm1.4$    \\ 
              $ g_{\chi_{c2} \chi_{c2}}/g_{direct} $  & $-15.3\pm6.1$  & $-12.3\pm1.4$    \\ 
              $\mathcal{C}_{J/\psi J/\psi}$  & $-144\pm14$  & $-82.3\pm1.8$    \\
              $\mathcal{C}_{\eta_c\chi_{c1}}$  & $342\pm107$  & $-21.3\pm18.7$    \\ 
              $\mathcal{C}_{J/\psi\psi(3686)}$  & $-20\pm40$  & $-32.0\pm5.5$    \\ 
              $\mathcal{C}_{\chi_{c0}\chi_{c1}}$  & $380\pm54$  & $20.0\pm50.6$    \\ 
              $\mathcal{C}_{\chi_{c2} \chi_{c2}}$  & $145\pm175$  & $-41.4\pm13.3$    \\ 
               $\alpha $ ~(GeV)  & $0.871\pm0.046$  & $1.813\pm0.030$    \\                  
                     \bottomrule[0.6pt]     
                      $\chi^2/d.o.f.$ & 0.657  & 0.699    \\  
			\bottomrule[1.0pt]
		\end{tabular}\label{table:parameter}}
  \end{table}

Beforehand, we further perform an analysis of the fit-II scheme to CMS's experimental data by assuming that the coupling strengths $\mathcal{C}$ for three channels of $\chi_{c1} \eta_c$, $\chi_{c0} \chi_{c1}$, and $\chi_{c2} \chi_{c2}$ are not powerful enough to generate  resonance poles in the dynamical mechanism. The corresponding best-fitting results are presented in Fig. \ref{fig:Fitexp} (a) with a dashed green line, where a $\chi^2/d.o.f.$ value is 0.699. It can be seen that the fitted  $\chi^2$ value in the fit-II scheme is slightly larger than that in the  fit I scheme, but there is no essential difference between two fitting line shapes of the di-$J/\psi$ mass spectrum, where several enhancement or dip structures in the di-$J/\psi$ spectrum can be reproduced in both the fit-I and fit-II scenarios.  By a pole analysis, we found that there are not any pole structures in the fit-II scheme and the corresponding central values of the coupling constants $\mathcal{C}_{\chi_{c1} \eta_c}=-21.3$, $\mathcal{C}_{J/\psi \psi(3686)}=-32.0$,  $\mathcal{C}_{\chi_{c0} \chi_{c1}}=20.0$ and $\mathcal{C}_{\chi_{c2} \chi_{c2}}=-41.4$ with cutoff $\alpha=1.813$ shown in Table \ref{table:parameter} will induce a threshold cusp line shape on the double $J/\psi$ spectrum.  This finding actually means that the threshold cusp or resonance pole solution cannot be definitely distinguished from the present experimental precision,  which should depend on the concrete coupling strength of double charmonia scattering.

{ In addition to the resonance poles, ignoring small width effect from intermediate charmonium states, we found several  virtual poles below the respective threshold in the two fit schemes, which are 
\begin{eqnarray}
\mathcal{E}_{J/\psi J/\psi}^{\mathrm{I}}&=&6.191~~\mathrm{GeV}, ~~~\mathcal{E}_{J/\psi \psi(2S)}^{\mathrm{I}}=6.718 ~~\mathrm{GeV}, \nonumber \\
\mathcal{E}_{J/\psi J/\psi}^{\mathrm{II}}&=&6.188~~\mathrm{GeV}, ~~~\mathcal{E}_{J/\psi \psi(2S)}^{\mathrm{II}}=6.687~~\mathrm{GeV}, \\
\mathcal{E}_{\eta_c \chi_{c1}}^{\mathrm{II}}&=&6.343~~\mathrm{GeV}, ~~~\mathcal{E}_{\chi_{c2}\chi_{c2}}^{\mathrm{II}}=7.045~~\mathrm{GeV}, \nonumber 
\end{eqnarray}
where superscript I and II represent the fit-I and fit-II scheme, respectively. It can be seen that these virtual pole effects produce the obvious threshold cusp structures.  }

{ We also studied the scattering lengths of each intermediate charmonium channel in the two fit schemes. The interaction property of double charmonia scattering can be reflected by scattering length $a_0$; i.e., positive $a_0$ and negative $a_0$ correspond to attractive and repulsive interaction in the absence of a bound state pole, respectively \cite{Dong:2020hxe}. By  effective range expansion, the scattering amplitude in the immediate vicinity of the threshold can be written as 
\begin{equation}
A_0^{-1}=\frac{1}{a_0}-i\sqrt{2\mu E},
\end{equation}
where $\mu=m_i m_j/(m_i+m_j)$ and $E$ is the energy relative to the two-body threshold. Obviously, the scattering length can be obtained by $a_0=A_0(E)|_{E\to 0}$. Without considering the width effect from intermediate charmonium states, we calculated the scattering length $a_0({ij})$ for each scattering channel and summarized them in Table \ref{table:a0}. It can be seen that their scattering lengths imply an attractive interaction, which is not contradictory with the produced threshold cusps or pole structures. }

\begin{figure}[t]
\includegraphics[width=8cm,keepaspectratio]{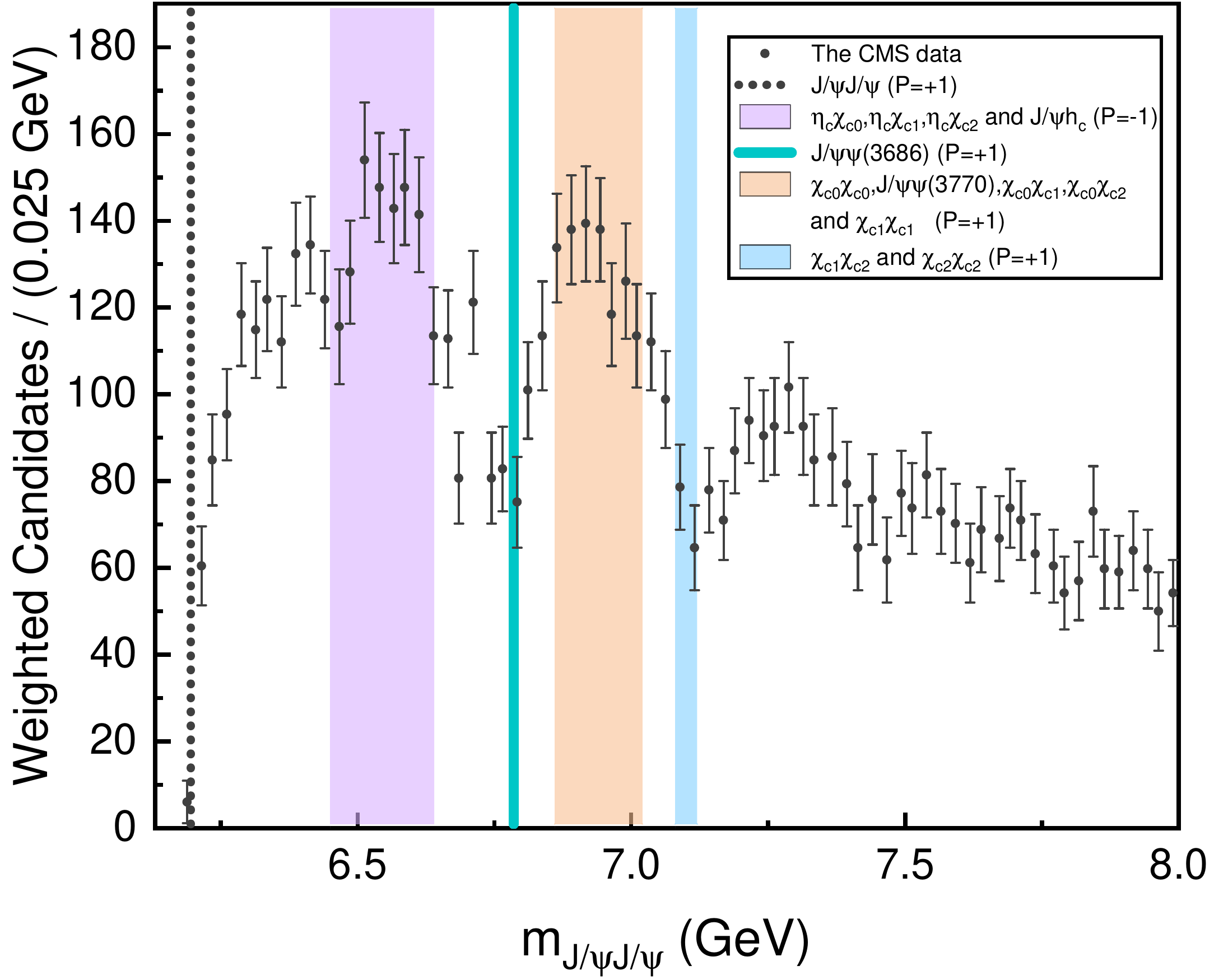}
	\caption{ %The comparison between the peak positions of the predicted compact fully-charm tetraquark  state \cite{Wang:2019rdo,Wang:2021kfv} and the peaking structures generated from the dynamical mechanism (left panel).  
	The comparison between the CMS data \cite{Cms2022} and the distribution of the threshold position of all of allowed intermediate charmonium pairs. \label{fig:comp2}  }
\end{figure}

Anyway,  it can be found that the threshold positions of all allowed combinations of intermediate charmonium pairs in the di-$J/\psi$ energy region from 6.20 to 7.30 GeV can be assigned to four main regions, which are  ($6.45-6.64$) GeV, 6.783 GeV, ($6.87-7.00$) GeV, and ($7.03-7.13$) GeV. Very interestingly, we can notice that two observed enhancements and two dips in CMS's measurement data exactly correspond to the above four characteristic energy regions in sequence as shown in Fig. \ref{fig:comp2}. This perfect agreement should provide a very strong hint to support our proposed dynamical mechanism for explaining these novel fully charm peaking structures in the  di-$J/\psi$ mass spectrum. We expect that this mechanism can be confirmed in future precise experimental measurements, especially at run III of the LHC.

\begin{table}[ht]
  	\caption{The calculated center value of the scattering length for each intermediate channel by using parameters in the fit-I and fit-II schemes.}
  	\setlength{\tabcolsep}{7mm}{
  	\begin{tabular}{lccccccc}
			\toprule[1.0pt] 
     Scattering length  &  Fit I  &  Fit II  \\
			\toprule[0.6pt] 
              $a_0({J/\psi J/\psi})$ (fm)  & $2.25$  & $1.64$    \\
              $a_0({\eta_c\chi_{c1}})$ (fm)  & $0.33$  & $0.48$    \\ 
              $a_0({J/\psi\psi(3686)})$ (fm)  & $0.87$  & $0.53$    \\ 
              $a_0({\chi_{c0}\chi_{c1}})$  (fm) & $0.32$  & $0.33$    \\ 
              $a_0({\chi_{c2} \chi_{c2}})$ (fm)  & $0.51$  & $0.58$    \\                
                     \bottomrule[0.6pt]    
			\bottomrule[1.0pt]
		\end{tabular}\label{table:a0}}
  \end{table}

\section{Discussion and Conclusion}\label{sec4}

%{\noindent\it Conclusion.}---

Very recently, the CMS Collaboration reported  measurement results of the invariant mass spectrum of the $J/\psi$ pair from $pp$ collisions, where several peaking phenomena were observed \cite{Cms2022}. Compared with previously reported $X(6900)$ structure by LHCb \cite{LHCb:2020bwg}, the new CMS measurement brings us more  important information. In this work, we have studied these newly observed fully charm enhancement structures in an extended dynamical mechanism. Our basic idea is based on a reaction that different combinations of the intermediate double charmonia directly produced in high-energy proton-proton collisions are transferred into final states of $J/\psi J/\psi$.  This reaction picture can be further extended to the contribution involving higher-order loops.

By employing the extended dynamical model to describe the line shape of the invariant mass spectrum of double $J/\psi$ newly measured by CMS, we have demonstrated that the contributions of all four characteristic intermediate channels $\eta_c\chi_{c1}$, $J/\psi \psi(3686)$, $\chi_{c0}\chi_{c1}$, and $\chi_{c2} \chi_{c2}$ are required in order to reproduce the CMS distribution. This fact means that these new features from the CMS measurement provide strong evidence to support the dynamical interpretation for the observed fully charm enhancement structures. Furthermore, we adopted two fitting schemes to explore the origin of the fully charm peaking phenomena in the dynamical mechanism, in which we concluded that the threshold cusp and resonance pole solution  cannot be distinguished from the present experimental precision.

%In order to better explore the inner interaction of the novel double charmonia scattering in our proposed dynamical production mechanism, we suggest two accessible ways here.

 \begin{figure}[t]
\includegraphics[width=8cm,keepaspectratio]{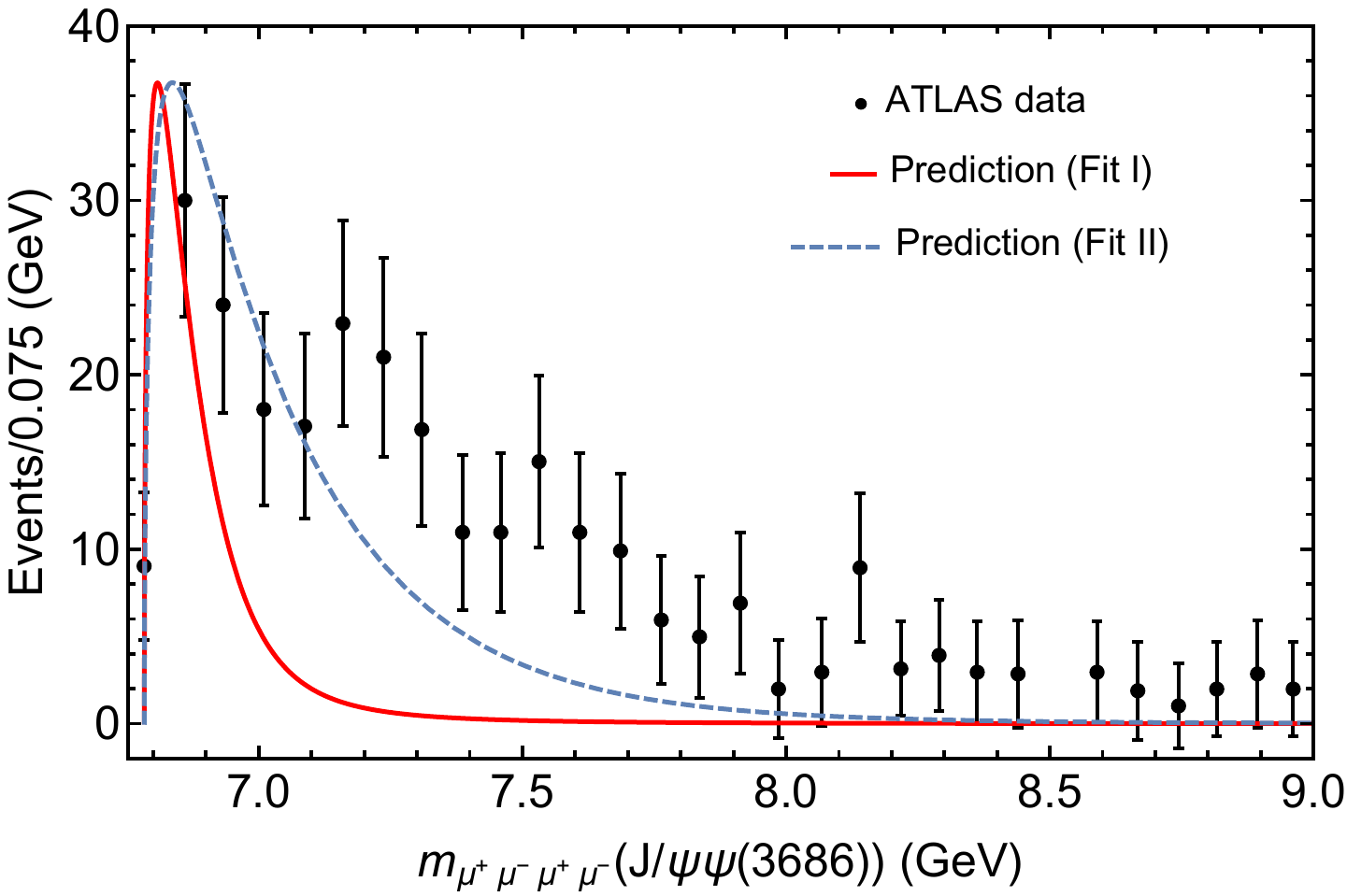}
	\caption{ The comparison between the ATLAS data and our predictions for the line shape of the invariant mass spectrum of $J/\psi \psi(3686)$ from the $J/\psi \psi(3686)$ channel in two parameter schemes. \label{fig:atlas}  }
\end{figure}

In order to better solve this problem, we suggest two accessible ways here. One can find there exist many combinations of on-shell charmonium pairs in the di-$J/\psi$ mass spectrum because of an approximate heavy quark symmetry in the charm sector, whose interference effect usually causes  difficulty of identifying the origin of these novel fully charm enhancements. Thus, an available method is to measure the invariant mass spectrum of the intermediate channel itself in hadron colliders, such as $\eta_c \chi_{cJ}$, $J/\psi\psi(3686)$, $J/\psi\psi(3770)$,  $\chi_{cJ}\chi_{cJ}$, and so on, where the line shape measurement should be helpful to identify the threshold cusp or resonance solution because the contributions of some off-shell channels should be suppressed by the phase space.  { We noticed that the ATLAS Collaboration recently released the preliminary result of the $J/\psi \psi(3686)$ mass spectrum,  which just can  test our dynamical production mechanism of double charmonia. In Fig. \ref{fig:atlas}, we presented the comparison between the ATLAS data and our predictions for the line shape of the invariant mass spectrum of $J/\psi \psi(3686)$ from the $J/\psi \psi(3686)$ channel in two parameter schemes. It can be seen that the obvious enhancement near threshold can be explained well, which further supports our theoretical mechanism. We believe that more precise experimental data of the $J/\psi \psi(3686)$ mass spectrum is helpful for definitely determining  the coupling behavior of $J/\psi \psi(3686)$ channel.  }

The second approach is to quantitatively estimate the magnitude of coupling strength of double charmonia scattering  by the first principle lattice QCD theory. Coincidentally, we noticed recent research work to discuss the property of a dibaryon scattering of $\Omega_{ccc}^{++}\Omega_{ccc}^{++}$  near unitarity region from lattice QCD \cite{Lyu:2021qsh}, which is another novel fully charm system. We hope that our analysis in this work can stimulate activities from lattice group to study double charmonia scattering, which should be worth expecting in the future.

Finally, we strongly call for more theoretical and experimental studies to concentrate on this kind of novel fully heavy system, which should be a new important frontier to investigate nonperturbative behavior of strong interaction in the future.

\section*{ACKNOWLEDGEMENTS}

This work is supported by the China National Funds for Distinguished Young Scientists under Grant No. 11825503, National Key Research and Development Program of China under Contract No. 2020YFA0406400, the 111 Project under Grant No. B20063, and the National Natural Science Foundation of China under Grant No. 12047501. J.-Z. W. is also supported by the National Postdoctoral Program for Innovative Talent.


\begin{thebibliography}{199}


%\cite{LHCb:2019kea}
\bibitem{LHCb:2019kea}
R.~Aaij \textit{et al.} [LHCb],
Observation of a narrow pentaquark state, $P_c(4312)^+$, and of two-peak structure of the $P_c(4450)^+$,
Phys. Rev. Lett. \textbf{122}, no.22, 222001 (2019).
%doi:10.1103/PhysRevLett.122.222001
%[arXiv:1904.03947 [hep-ex]].
%467 citations counted in INSPIRE as of 10 Jul 2022


%\cite{Belle:2003nnu}
\bibitem{Belle:2003nnu}
S.~K.~Choi \textit{et al.} [Belle],
Observation of a narrow charmonium-like state in exclusive $B^\pm \to K^\pm \pi^+ \pi^- J/\psi$ decays,
Phys. Rev. Lett. \textbf{91}, 262001 (2003).
%doi:10.1103/PhysRevLett.91.262001
%[arXiv:hep-ex/0309032 [hep-ex]].
%2156 citations counted in INSPIRE as of 11 Jul 2022


%\cite{BESIII:2013ris}
\bibitem{BESIII:2013ris}
M.~Ablikim \textit{et al.} [BESIII],
Observation of a Charged Charmoniumlike Structure in $e^+e^- \to \pi^+\pi^- J/\psi$ at $\sqrt{s}$ =4.26  GeV,
Phys. Rev. Lett. \textbf{110}, 252001 (2013).
%doi:10.1103/PhysRevLett.110.252001
%[arXiv:1303.5949 [hep-ex]].
%948 citations counted in INSPIRE as of 10 Jul 2022



%\cite{BESIII:2013ouc}
\bibitem{BESIII:2013ouc}
M.~Ablikim \textit{et al.} [BESIII],
Observation of a Charged Charmoniumlike Structure $Z_c$(4020) and Search for the $Z_c$(3900) in $e^+e^- \to \pi^+\pi^-h_c$,
Phys. Rev. Lett. \textbf{111}, no.24, 242001 (2013).
%doi:10.1103/PhysRevLett.111.242001
%[arXiv:1309.1896 [hep-ex]].
%450 citations counted in INSPIRE as of 10 Jul 2022


%\cite{LHCb:2021uow}
\bibitem{LHCb:2021uow}
R.~Aaij \textit{et al.} [LHCb],
Observation of New Resonances Decaying to $J/\psi K^+$ and $J/\psi \phi$,
Phys. Rev. Lett. \textbf{127}, no.8, 082001 (2021).
%doi:10.1103/PhysRevLett.127.082001
%[arXiv:2103.01803 [hep-ex]].
%84 citations counted in INSPIRE as of 11 Jul 2022



%\cite{LHCb:2021auc}
\bibitem{LHCb:2021auc}
R.~Aaij \textit{et al.} [LHCb],
Study of the doubly charmed tetraquark $T_{cc}^{+}$,
Nature Commun. \textbf{13}, no.1, 3351 (2022).
%doi:10.1038/s41467-022-30206-w
%[arXiv:2109.01056 [hep-ex]].
%107 citations counted in INSPIRE as of 10 Jul 2022



%\cite{Chen:2016qju}
\bibitem{Chen:2016qju}
  H.~X.~Chen, W.~Chen, X.~Liu and S.~L.~Zhu,
  The hidden-charm pentaquark and tetraquark states,
  Phys.\ Rept.\  {\bf 639} (2016) 1.
%  doi:10.1016/j.physrep.2016.05.004
%  [arXiv:1601.02092 [hep-ph]].
  %%CITATION = doi:10.1016/j.physrep.2016.05.004;%%
  %499 citations counted in INSPIRE as of 08 Feb 2020

%\cite{Liu:2019zoy}
\bibitem{Liu:2019zoy}
  Y.~R.~Liu, H.~X.~Chen, W.~Chen, X.~Liu and S.~L.~Zhu,
  Pentaquark and Tetraquark states,
  Prog.\ Part.\ Nucl.\ Phys.\  {\bf 107}, 237 (2019).
%  doi:10.1016/j.ppnp.2019.04.003
%  [arXiv:1903.11976 [hep-ph]].
  %%CITATION = doi:10.1016/j.ppnp.2019.04.003;%%
  %77 citations counted in INSPIRE as of 08 Feb 2020
  
  
%\cite{Chen:2022asf}
\bibitem{Chen:2022asf}
H.~X.~Chen, W.~Chen, X.~Liu, Y.~R.~Liu and S.~L.~Zhu,
An updated review of the new hadron states,
[arXiv:2204.02649 [hep-ph]].
%31 citations counted in INSPIRE as of 10 Jul 2022
  
  
  %\cite{Guo:2017jvc}
\bibitem{Guo:2017jvc}
F.~K.~Guo, C.~Hanhart, U.~G.~Meißner, Q.~Wang, Q.~Zhao and B.~S.~Zou,
Hadronic molecules,
Rev. Mod. Phys. \textbf{90}, no.1, 015004 (2018).
%doi:10.1103/RevModPhys.90.015004
%[arXiv:1705.00141 [hep-ph]].
%404 citations counted in INSPIRE as of 07 Jul 2020


%\cite{Olsen:2017bmm}
\bibitem{Olsen:2017bmm}
S.~L.~Olsen, T.~Skwarnicki and D.~Zieminska,
Nonstandard heavy mesons and baryons: Experimental evidence,
Rev. Mod. Phys. \textbf{90}, no.1, 015003 (2018).
%doi:10.1103/RevModPhys.90.015003
%[arXiv:1708.04012 [hep-ph]].
%259 citations counted in INSPIRE as of 09 Jul 2020



%\cite{Brambilla:2019esw}
\bibitem{Brambilla:2019esw}
N.~Brambilla, S.~Eidelman, C.~Hanhart, A.~Nefediev, C.~P.~Shen, C.~E.~Thomas, A.~Vairo and C.~Z.~Yuan,
The $XYZ$ states: experimental and theoretical status and perspectives,
Phys. Rept. \textbf{873}, 1-154 (2020).
%doi:10.1016/j.physrep.2020.05.001
%[arXiv:1907.07583 [hep-ex]].
%369 citations counted in INSPIRE as of 10 Jul 2022


%\cite{LHCb:2020bwg}
\bibitem{LHCb:2020bwg}
R.~Aaij \textit{et al.} [LHCb],
Observation of structure in the $J /\psi$ -pair mass spectrum,
Sci. Bull. \textbf{65}, no.23, 1983-1993 (2020).
%doi:10.1016/j.scib.2020.08.032
%[arXiv:2006.16957 [hep-ex]].
%190 citations counted in INSPIRE as of 10 Jul 2022





\bibitem{Atlas2022}
Evelina Bouhova-Thacker on behalf of the ATLAS Collaboration,
ATLAS results on exotic hadronic resonances, \href{https://cds.cern.ch/record/2819987}{Proceedings at ICHEP 2022}, 
https://agenda.infn.it/event/28874/contributions/170298/.



%\cite{Chen:2020xwe}
\bibitem{Chen:2020xwe}
H.~X.~Chen, W.~Chen, X.~Liu and S.~L.~Zhu,
Strong decays of fully-charm tetraquarks into di-charmonia,
Sci. Bull. \textbf{65}, 1994-2000 (2020).
%doi:10.1016/j.scib.2020.08.038
%[arXiv:2006.16027 [hep-ph]].
%51 citations counted in INSPIRE as of 10 Jul 2022


%\cite{Jin:2020jfc}
\bibitem{Jin:2020jfc}
X.~Jin, Y.~Xue, H.~Huang and J.~Ping,
Full-heavy tetraquarks in constituent quark models,
Eur. Phys. J. C \textbf{80}, no.11, 1083 (2020).
%doi:10.1140/epjc/s10052-020-08650-z
%[arXiv:2006.13745 [hep-ph]].
%48 citations counted in INSPIRE as of 10 Jul 2022


%\cite{Lu:2020cns}
\bibitem{Lu:2020cns}
Q.~F.~L\"u, D.~Y.~Chen and Y.~B.~Dong,
Masses of fully heavy tetraquarks $QQ {\bar{Q}} {\bar{Q}}$ in an extended relativized quark model,
Eur. Phys. J. C \textbf{80}, no.9, 871 (2020).
%doi:10.1140/epjc/s10052-020-08454-1
%[arXiv:2006.14445 [hep-ph]].
%61 citations counted in INSPIRE as of 10 Jul 2022




%\cite{Yang:2020rih}
\bibitem{Yang:2020rih}
G.~Yang, J.~Ping, L.~He and Q.~Wang,
A potential model prediction of fully-heavy tetraquarks $QQ\bar{Q}\bar{Q}$ ($Q=c, b$),
[arXiv:2006.13756 [hep-ph]].
%3 citations counted in INSPIRE as of 29 Jul 2020



%\cite{Deng:2020iqw}
\bibitem{Deng:2020iqw}
C.~Deng, H.~Chen and J.~Ping,
Towards the understanding of fully-heavy tetraquark states from various models,
Phys. Rev. D \textbf{103}, no.1, 014001 (2021).
%doi:10.1103/PhysRevD.103.014001
%[arXiv:2003.05154 [hep-ph]].
%46 citations counted in INSPIRE as of 10 Jul 2022



%\cite{Wang:2020ols}
\bibitem{Wang:2020ols}
Z.~G.~Wang,
Tetraquark candidates in the LHCb's di-$J/\psi$ mass spectrum,
Chin. Phys. C \textbf{44}, no.11, 113106 (2020).
%doi:10.1088/1674-1137/abb080
%[arXiv:2006.13028 [hep-ph]].
%39 citations counted in INSPIRE as of 10 Jul 2022



%\cite{Chen:2020lgj}
\bibitem{Chen:2020lgj}
X.~Chen,
Fully-charm tetraquarks: $cc\bar{c}\bar{c}$,
[arXiv:2001.06755 [hep-ph]].
%4 citations counted in INSPIRE as of 29 Jul 2020



%\cite{Albuquerque:2020hio}
\bibitem{Albuquerque:2020hio}
R.~M.~Albuquerque, S.~Narison, A.~Rabemananjara, D.~Rabetiarivony and G.~Randriamanatrika,
Doubly-hidden scalar heavy molecules and tetraquarks states from QCD at NLO,
Phys. Rev. D \textbf{102}, no.9, 094001 (2020).
%doi:10.1103/PhysRevD.102.094001
%[arXiv:2008.01569 [hep-ph]].
%43 citations counted in INSPIRE as of 10 Jul 2022



%\cite{Sonnenschein:2020nwn}
\bibitem{Sonnenschein:2020nwn}
J.~Sonnenschein and D.~Weissman,
Deciphering the recently discovered tetraquark candidates around 6.9 GeV,
Eur. Phys. J. C \textbf{81}, no.1, 25 (2021).
%doi:10.1140/epjc/s10052-020-08818-7
%[arXiv:2008.01095 [hep-ph]].
%32 citations counted in INSPIRE as of 10 Jul 2022



%\cite{Giron:2020wpx}
\bibitem{Giron:2020wpx}
J.~F.~Giron and R.~F.~Lebed,
Simple spectrum of $c\bar c c\bar c$ states in the dynamical diquark model,
Phys. Rev. D \textbf{102}, no.7, 074003 (2020).
%doi:10.1103/PhysRevD.102.074003
%[arXiv:2008.01631 [hep-ph]].
%55 citations counted in INSPIRE as of 10 Jul 2022




%\cite{Richard:2020hdw}
\bibitem{Richard:2020hdw}
J.~M.~Richard,
About the $J/\psi$ $J/\psi$ peak of LHCb: fully-charmed tetraquark?,
Sci. Bull. \textbf{65}, 1954-1955 (2020).
%doi:10.1016/j.scib.2020.08.020
%[arXiv:2008.01962 [hep-ph]].
%30 citations counted in INSPIRE as of 10 Jul 2022


%\cite{Becchi:2020uvq}
\bibitem{Becchi:2020uvq}
C.~Becchi, J.~Ferretti, A.~Giachino, L.~Maiani and E.~Santopinto,
A study of $c c\bar{c}\bar{c}$ tetraquark decays in 4 muons and in $D^{(*)} \bar{D}^{(*)}$ at LHC,
Phys. Lett. B \textbf{811}, 135952 (2020).
%doi:10.1016/j.physletb.2020.135952
%[arXiv:2006.14388 [hep-ph]].
%37 citations counted in INSPIRE as of 10 Jul 2022


%\cite{liu:2020eha}
\bibitem{liu:2020eha}
M.~S.~Liu, F.~X.~Liu, X.~H.~Zhong and Q.~Zhao,
Full-heavy tetraquark states and their evidences in the LHCb di-$J/\psi$ spectrum,
[arXiv:2006.11952 [hep-ph]].
%7 citations counted in INSPIRE as of 28 Aug 2020



%\cite{Bedolla:2019zwg}
\bibitem{Bedolla:2019zwg}
M.~A.~Bedolla, J.~Ferretti, C.~D.~Roberts and E.~Santopinto,
Spectrum of fully-heavy tetraquarks from a diquark+antidiquark perspective,
Eur. Phys. J. C \textbf{80}, no.11, 1004 (2020).
%doi:10.1140/epjc/s10052-020-08579-3
%[arXiv:1911.00960 [hep-ph]].
%70 citations counted in INSPIRE as of 10 Jul 2022


%\cite{Chao:2020dml}
\bibitem{Chao:2020dml}
K.~T.~Chao and S.~L.~Zhu,
The possible tetraquark states $cc \bar c \bar c$ observed by the LHCb experiment,
Sci. Bull. \textbf{65}, no.23, 1952-1953 (2020).
%doi:10.1016/j.scib.2020.08.031
%[arXiv:2008.07670 [hep-ph]].
%31 citations counted in INSPIRE as of 10 Jul 2022


%\cite{Karliner:2020dta}
\bibitem{Karliner:2020dta}
M.~Karliner and J.~L.~Rosner,
Interpretation of structure in the di- $J/\psi$ spectrum,
Phys. Rev. D \textbf{102}, no.11, 114039 (2020).
%doi:10.1103/PhysRevD.102.114039
%[arXiv:2009.04429 [hep-ph]].
%47 citations counted in INSPIRE as of 10 Jul 2022



%\cite{Faustov:2020qfm}
\bibitem{Faustov:2020qfm}
R.~N.~Faustov, V.~O.~Galkin and E.~M.~Savchenko,
Masses of the $QQ\bar Q\bar Q$ tetraquarks in the relativistic diquark--antidiquark picture,
Phys. Rev. D \textbf{102}, no.11, 114030 (2020).
%doi:10.1103/PhysRevD.102.114030
%[arXiv:2009.13237 [hep-ph]].
%27 citations counted in INSPIRE as of 10 Jul 2022


%\cite{Gordillo:2020sgc}
\bibitem{Gordillo:2020sgc}
M.~C.~Gordillo, F.~De Soto and J.~Segovia,
Diffusion Monte Carlo calculations of fully-heavy multiquark bound states,
Phys. Rev. D \textbf{102}, no.11, 114007 (2020).
%doi:10.1103/PhysRevD.102.114007
%[arXiv:2009.11889 [hep-ph]].
%26 citations counted in INSPIRE as of 10 Jul 2022


%\cite{Weng:2020jao}
\bibitem{Weng:2020jao}
X.~Z.~Weng, X.~L.~Chen, W.~Z.~Deng and S.~L.~Zhu,
Systematics of fully heavy tetraquarks,
Phys. Rev. D \textbf{103}, no.3, 034001 (2021).
%doi:10.1103/PhysRevD.103.034001
%[arXiv:2010.05163 [hep-ph]].
%35 citations counted in INSPIRE as of 10 Jul 2022


%\cite{Zhang:2020xtb}
\bibitem{Zhang:2020xtb}
J.~R.~Zhang,
$0^{+}$ fully-charmed tetraquark states,
Phys. Rev. D \textbf{103}, no.1, 014018 (2021).
%doi:10.1103/PhysRevD.103.014018
%[arXiv:2010.07719 [hep-ph]].
%27 citations counted in INSPIRE as of 10 Jul 2022


%\cite{Yang:2020wkh}
\bibitem{Yang:2020wkh}
B.~C.~Yang, L.~Tang and C.~F.~Qiao,
Scalar fully-heavy tetraquark states $QQ^\prime {\bar{Q}} \bar{Q^\prime }$ in QCD sum rules,
Eur. Phys. J. C \textbf{81}, no.4, 324 (2021).
%doi:10.1140/epjc/s10052-021-09096-7
%[arXiv:2012.04463 [hep-ph]].
%21 citations counted in INSPIRE as of 10 Jul 2022



%\cite{Zhao:2020zjh}
\bibitem{Zhao:2020zjh}
Z.~Zhao, K.~Xu, A.~Kaewsnod, X.~Liu, A.~Limphirat and Y.~Yan,
Study of charmoniumlike and fully-charm tetraquark spectroscopy,
Phys. Rev. D \textbf{103}, no.11, 116027 (2021).
%doi:10.1103/PhysRevD.103.116027
%[arXiv:2012.15554 [hep-ph]].
%20 citations counted in INSPIRE as of 10 Jul 2022


%\cite{Faustov:2021hjs}
\bibitem{Faustov:2021hjs}
R.~N.~Faustov, V.~O.~Galkin and E.~M.~Savchenko,
Heavy tetraquarks in the relativistic quark model,
Universe \textbf{7}, no.4, 94 (2021).
%doi:10.3390/universe7040094
%[arXiv:2103.01763 [hep-ph]].
%33 citations counted in INSPIRE as of 10 Jul 2022


%\cite{Ke:2021iyh}
\bibitem{Ke:2021iyh}
H.~W.~Ke, X.~Han, X.~H.~Liu and Y.~L.~Shi,
Tetraquark state $X(6900)$ and the interaction between diquark and antidiquark,
Eur. Phys. J. C \textbf{81}, no.5, 427 (2021).
%doi:10.1140/epjc/s10052-021-09229-y
%[arXiv:2103.13140 [hep-ph]].
%18 citations counted in INSPIRE as of 10 Jul 2022


%\cite{Yang:2021hrb}
\bibitem{Yang:2021hrb}
G.~Yang, J.~Ping and J.~Segovia,
Exotic resonances of fully-heavy tetraquarks in a lattice-QCD insipired quark model,
Phys. Rev. D \textbf{104}, no.1, 014006 (2021).
%doi:10.1103/PhysRevD.104.014006
%[arXiv:2104.08814 [hep-ph]].
%13 citations counted in INSPIRE as of 10 Jul 2022


%\cite{Li:2021ygk}
\bibitem{Li:2021ygk}
Q.~Li, C.~H.~Chang, G.~L.~Wang and T.~Wang,
Mass spectra and wave functions of $T_{QQ\bar{Q}\bar{Q}}$ tetraquarks,
Phys. Rev. D \textbf{104}, no.1, 014018 (2021).
%doi:10.1103/PhysRevD.104.014018
%[arXiv:2104.12372 [hep-ph]].
%14 citations counted in INSPIRE as of 10 Jul 2022


%\cite{Asadi:2021ids}
\bibitem{Asadi:2021ids}
Z.~Asadi and G.~R.~Boroun,
Masses of fully heavy tetraquark states from a four-quark static potential model,
Phys. Rev. D \textbf{105}, no.1, 014006 (2022).
%doi:10.1103/PhysRevD.105.014006
%[arXiv:2112.11028 [hep-ph]].
%1 citations counted in INSPIRE as of 10 Jul 2022


%\cite{Kuang:2022vdy}
\bibitem{Kuang:2022vdy}
Z.~Kuang, K.~Serafin, X.~Zhao and J.~P.~Vary,
All-charm tetraquark in front form dynamics,
Phys. Rev. D \textbf{105}, 094028 (2022).
%doi:10.1103/PhysRevD.105.094028
%[arXiv:2201.06428 [hep-ph]].
%1 citations counted in INSPIRE as of 10 Jul 2022


%\cite{Wu:2022qwd}
\bibitem{Wu:2022qwd}
R.~H.~Wu, Y.~S.~Zuo, C.~Y.~Wang, C.~Meng, Y.~Q.~Ma and K.~T.~Chao,
NLO results with operator mixing for fully heavy tetraquarks in QCD sum rules,
[arXiv:2201.11714 [hep-ph]].
%2 citations counted in INSPIRE as of 10 Jul 2022


%\cite{Wang:2021kfv}
\bibitem{Wang:2021kfv}
G.~J.~Wang, L.~Meng, M.~Oka and S.~L.~Zhu,
Higher fully charmed tetraquarks: Radial excitations and P-wave states,
Phys. Rev. D \textbf{104}, no.3, 036016 (2021).
%doi:10.1103/PhysRevD.104.036016
%[arXiv:2105.13109 [hep-ph]].
%11 citations counted in INSPIRE as of 10 Jul 2022


%------------------------------------------------------------------------------------


%\cite{Wang:2020wrp}
\bibitem{Wang:2020wrp}
J.~Z.~Wang, D.~Y.~Chen, X.~Liu and T.~Matsuki,
Producing fully charm structures in the $J/\psi$ -pair invariant mass spectrum,
Phys. Rev. D \textbf{103}, no.7, 071503 (2021).
%doi:10.1103/PhysRevD.103.L071503
%[arXiv:2008.07430 [hep-ph]].
%34 citations counted in INSPIRE as of 10 Jul 2022


%\cite{Dong:2020nwy}
\bibitem{Dong:2020nwy}
X.~K.~Dong, V.~Baru, F.~K.~Guo, C.~Hanhart and A.~Nefediev,
Coupled-Channel Interpretation of the LHCb Double-~$J/\psi$~Spectrum and Hints of a New State Near the~ $J/\psi J/\psi$~~Threshold,
Phys. Rev. Lett. \textbf{126}, no.13, 132001 (2021)
[erratum: Phys. Rev. Lett. \textbf{127}, no.11, 119901 (2021)].
%doi:10.1103/PhysRevLett.127.119901
%[arXiv:2009.07795 [hep-ph]].
%45 citations counted in INSPIRE as of 10 Jul 2022



%\cite{Guo:2020pvt}
\bibitem{Guo:2020pvt}
Z.~H.~Guo and J.~A.~Oller,
Insights into the inner structures of the fully charmed tetraquark state $X(6900)$,
Phys. Rev. D \textbf{103}, no.3, 034024 (2021).
%doi:10.1103/PhysRevD.103.034024
%[arXiv:2011.00978 [hep-ph]].
%29 citations counted in INSPIRE as of 10 Jul 2022


%\cite{Liang:2021fzr}
\bibitem{Liang:2021fzr}
Z.~R.~Liang, X.~Y.~Wu and D.~L.~Yao,
Hunting for states in the recent LHCb di-J/\ensuremath{\psi} invariant mass spectrum,
Phys. Rev. D \textbf{104}, no.3, 034034 (2021).
%doi:10.1103/PhysRevD.104.034034
%[arXiv:2104.08589 [hep-ph]].
%15 citations counted in INSPIRE as of 10 Jul 2022


%\cite{Dong:2021lkh}
\bibitem{Dong:2021lkh}
X.~K.~Dong, V.~Baru, F.~K.~Guo, C.~Hanhart, A.~Nefediev and B.~S.~Zou,
Is the existence of a J/\ensuremath{\psi}J/\ensuremath{\psi} bound state plausible?,
Sci. Bull. \textbf{66}, no.24, 2462-2470 (2021).
%doi:10.1016/j.scib.2021.09.009
%[arXiv:2107.03946 [hep-ph]].
%10 citations counted in INSPIRE as of 10 Jul 2022


%\cite{Zhuang:2021pci}
\bibitem{Zhuang:2021pci}
Z.~Zhuang, Y.~Zhang, Y.~Ma and Q.~Wang,
Lineshape of the compact fully heavy tetraquark,
Phys. Rev. D \textbf{105}, no.5, 054026 (2022).
%doi:10.1103/PhysRevD.105.054026
%[arXiv:2111.14028 [hep-ph]].
%2 citations counted in INSPIRE as of 10 Jul 2022



%\cite{Wan:2020fsk}
\bibitem{Wan:2020fsk}
B.~D.~Wan and C.~F.~Qiao,
Gluonic tetracharm configuration of $X (6900)$,
Phys. Lett. B \textbf{817}, 136339 (2021).
%doi:10.1016/j.physletb.2021.136339
%[arXiv:2012.00454 [hep-ph]].
%24 citations counted in INSPIRE as of 10 Jul 2022



%\cite{Zhu:2020snb}
\bibitem{Zhu:2020snb}
J.~W.~Zhu, X.~D.~Guo, R.~Y.~Zhang, W.~G.~Ma and X.~Q.~Li,
A possible interpretation for $X(6900)$ observed in four-muon final state by LHCb-A light Higgs-like boson?,
[arXiv:2011.07799 [hep-ph]].
%18 citations counted in INSPIRE as of 10 Jul 2022



%\cite{Maciula:2020wri}
\bibitem{Maciula:2020wri}
R.~Maciu\l{}a, W.~Sch\"afer and A.~Szczurek,
On the mechanism of $T_{4c}$(6900) tetraquark production,
Phys. Lett. B \textbf{812}, 136010 (2021).
%doi:10.1016/j.physletb.2020.136010
%[arXiv:2009.02100 [hep-ph]].
%34 citations counted in INSPIRE as of 10 Jul 2022



%\cite{Ma:2020kwb}
\bibitem{Ma:2020kwb}
Y.~Q.~Ma and H.~F.~Zhang,
Exploring the Di-$J/\psi$ Resonances around 6.9 $\mathrm{GeV}$ Based on $ab$ $initio$ Perturbative QCD,
[arXiv:2009.08376 [hep-ph]].
%28 citations counted in INSPIRE as of 10 Jul 2022


%\cite{Feng:2020riv}
\bibitem{Feng:2020riv}
F.~Feng, Y.~Huang, Y.~Jia, W.~L.~Sang, X.~Xiong and J.~Y.~Zhang,
Fragmentation production of fully-charmed tetraquarks at LHC,
[arXiv:2009.08450 [hep-ph]].
%22 citations counted in INSPIRE as of 10 Jul 2022



%\cite{Zhu:2020xni}
\bibitem{Zhu:2020xni}
R.~Zhu,
Fully-heavy tetraquark spectra and production at hadron colliders,
Nucl. Phys. B \textbf{966}, 115393 (2021).
%doi:10.1016/j.nuclphysb.2021.115393
%[arXiv:2010.09082 [hep-ph]].
%29 citations counted in INSPIRE as of 10 Jul 2022


%\cite{Feng:2020qee}
\bibitem{Feng:2020qee}
F.~Feng, Y.~Huang, Y.~Jia, W.~L.~Sang and J.~Y.~Zhang,
Exclusive radiative production of fully-charmed tetraquarks at $B$ Factory,
Phys. Lett. B \textbf{818}, 136368 (2021).
%doi:10.1016/j.physletb.2021.136368
%[arXiv:2011.03039 [hep-ph]].
%10 citations counted in INSPIRE as of 10 Jul 2022


%\cite{Gong:2020bmg}
\bibitem{Gong:2020bmg}
C.~Gong, M.~C.~Du, Q.~Zhao, X.~H.~Zhong and B.~Zhou,
Nature of $X(6900)$ and its production mechanism at LHCb,
Phys. Lett. B \textbf{824}, 136794 (2022).
%doi:10.1016/j.physletb.2021.136794
%[arXiv:2011.11374 [hep-ph]].
%22 citations counted in INSPIRE as of 10 Jul 2022


%\cite{Goncalves:2021ytq}
\bibitem{Goncalves:2021ytq}
V.~P.~Gon\c{c}alves and B.~D.~Moreira,
Fully-heavy tetraquark production by $\gamma\gamma$ interactions in hadronic collisions at the LHC,
Phys. Lett. B \textbf{816}, 136249 (2021).
%doi:10.1016/j.physletb.2021.136249
%[arXiv:2101.03798 [hep-ph]].
%11 citations counted in INSPIRE as of 10 Jul 2022


%------------------------------------------------------------------------------------



\bibitem{Cms2022}
Kai Yi on behalf of the CMS Collaboration,
Recent CMS results on exotic resonance, \href{https://cds.cern.ch/record/2815336}{Proceedings at ICHEP 2022}, 
https://agenda.infn.it/event/28874/contributions/170300/.



%\cite{Guo:2014iya}
\bibitem{Guo:2014iya}
F.~K.~Guo, C.~Hanhart, Q.~Wang and Q.~Zhao,
Could the near-threshold $XYZ$ states be simply kinematic effects?,
Phys. Rev. D \textbf{91}, no.5, 051504 (2015).
%doi:10.1103/PhysRevD.91.051504
%[arXiv:1411.5584 [hep-ph]].
%111 citations counted in INSPIRE as of 10 Jul 2022



%\cite{Sun:2014gca}
\bibitem{Sun:2014gca}
L.~P.~Sun, H.~Han and K.~T.~Chao,
Impact of $J/\psi$ pair production at the LHC and predictions in nonrelativistic QCD,
Phys. Rev. D \textbf{94}, no.7, 074033 (2016).
%doi:10.1103/PhysRevD.94.074033
%[arXiv:1404.4042 [hep-ph]].
%53 citations counted in INSPIRE as of 02 Aug 2020


%\cite{Likhoded:2016zmk}
\bibitem{Likhoded:2016zmk}
A.~K.~Likhoded, A.~V.~Luchinsky and S.~V.~Poslavsky,
Production of $J/\psi + \chi_c$ and $J/\psi + J/\psi$ with real gluon emission at LHC,
Phys. Rev. D \textbf{94}, no.5, 054017 (2016).
%doi:10.1103/PhysRevD.94.054017
%[arXiv:1606.06767 [hep-ph]].
%28 citations counted in INSPIRE as of 02 Aug 2020


%\cite{Baranov:2011zz}
\bibitem{Baranov:2011zz}
S.~P.~Baranov,
Pair production of $J/\psi$ mesons in the $k_t$-factorization approach,
Phys. Rev. D \textbf{84}, 054012 (2011).
%doi:10.1103/PhysRevD.84.054012
%29 citations counted in INSPIRE as of 02 Aug 2020


%\cite{Lansberg:2013qka}
\bibitem{Lansberg:2013qka}
J.~P.~Lansberg and H.~S.~Shao,
Production of $J/\psi + \eta_{c}$ versus $J/\psi + J/\psi$ at the LHC: Importance of Real $\alpha^{5}_{s}$ Corrections,
Phys. Rev. Lett. \textbf{111}, 122001 (2013).
%doi:10.1103/PhysRevLett.111.122001
%[arXiv:1308.0474 [hep-ph]].
%71 citations counted in INSPIRE as of 02 Aug 2020


%\cite{Lansberg:2014swa}
\bibitem{Lansberg:2014swa}
J.~P.~Lansberg and H.~S.~Shao,
$J/\psi$ -pair production at large momenta: Indications for double parton scatterings and large $\alpha_s^5$ contributions,
Phys. Lett. B \textbf{751}, 479-486 (2015).
%doi:10.1016/j.physletb.2015.10.083
%[arXiv:1410.8822 [hep-ph]].
%88 citations counted in INSPIRE as of 02 Aug 2020


%\cite{Lansberg:2015lva}
\bibitem{Lansberg:2015lva}
J.~P.~Lansberg and H.~S.~Shao,
Double-quarkonium production at a fixed-target experiment at the LHC (AFTER@LHC),
Nucl. Phys. B \textbf{900}, 273-294 (2015).
%doi:10.1016/j.nuclphysb.2015.09.005
%[arXiv:1504.06531 [hep-ph]].
%46 citations counted in INSPIRE as of 02 Aug 2020


%\cite{Shao:2012iz}
\bibitem{Shao:2012iz}
H.~S.~Shao,
HELAC-Onia: An automatic matrix element generator for heavy quarkonium physics,
Comput. Phys. Commun. \textbf{184}, 2562-2570 (2013).
%doi:10.1016/j.cpc.2013.05.023
%[arXiv:1212.5293 [hep-ph]].
%90 citations counted in INSPIRE as of 02 Aug 2020


%\cite{Shao:2015vga}
\bibitem{Shao:2015vga}
H.~S.~Shao,
HELAC-Onia 2.0: an upgraded matrix-element and event generator for heavy quarkonium physics,
Comput. Phys. Commun. \textbf{198}, 238-259 (2016).
%doi:10.1016/j.cpc.2015.09.011
%[arXiv:1507.03435 [hep-ph]].
%68 citations counted in INSPIRE as of 02 Aug 2020


%\cite{Calucci:1997ii}
\bibitem{Calucci:1997ii}
G.~Calucci and D.~Treleani,
Mini - jets and the two-body parton correlation,
Phys. Rev. D \textbf{57}, 503-511 (1998).
%doi:10.1103/PhysRevD.57.503
%[arXiv:hep-ph/9707389 [hep-ph]].
%37 citations counted in INSPIRE as of 02 Aug 2020


%\cite{Calucci:1999yz}
\bibitem{Calucci:1999yz}
G.~Calucci and D.~Treleani,
Proton structure in transverse space and the effective cross-section,
Phys. Rev. D \textbf{60}, 054023 (1999).
%doi:10.1103/PhysRevD.60.054023
%[arXiv:hep-ph/9902479 [hep-ph]].
%82 citations counted in INSPIRE as of 02 Aug 2020


%\cite{DelFabbro:2000ds}
\bibitem{DelFabbro:2000ds}
A.~Del Fabbro and D.~Treleani,
Scale factor in double parton collisions and parton densities in transverse space,
Phys. Rev. D \textbf{63}, 057901 (2001).
%doi:10.1103/PhysRevD.63.057901
%[arXiv:hep-ph/0005273 [hep-ph]].
%46 citations counted in INSPIRE as of 02 Aug 2020



%\cite{Lansberg:2019adr}
\bibitem{Lansberg:2019adr}
J.~P.~Lansberg,
New Observables in Inclusive Production of Quarkonia,
Phys. Rept. \textbf{889}, 1-106 (2020).
%doi:10.1016/j.physrep.2020.08.007
%[arXiv:1903.09185 [hep-ph]].
%56 citations counted in INSPIRE as of 28 Jan 2021



%\cite{Wang:2020tpt}
\bibitem{Wang:2020tpt}
J.~Z.~Wang, X.~Liu and T.~Matsuki,
Fully-heavy structures in the invariant mass spectrum of $J/\psi \psi(3686)$, $J/\psi \psi(3770)$, $\psi(3686) \psi(3686)$, and $J/\psi \Upsilon(1S)$ at hadron colliders,
Phys. Lett. B \textbf{816}, 136209 (2021).
%doi:10.1016/j.physletb.2021.136209
%[arXiv:2012.03281 [hep-ph]].
%8 citations counted in INSPIRE as of 10 Jul 2022



%\cite{He:2019qqr}
\bibitem{He:2019qqr}
Z.~G.~He, B.~A.~Kniehl, M.~A.~Nefedov and V.~A.~Saleev,
Double Prompt $J/\psi$ Hadroproduction in the Parton Reggeization Approach with High-Energy Resummation,
Phys. Rev. Lett. \textbf{123}, no.16, 162002 (2019).
%doi:10.1103/PhysRevLett.123.162002
%[arXiv:1906.08979 [hep-ph]].
%6 citations counted in INSPIRE as of 29 Jul 2020



%\cite{He:2015qya}
\bibitem{He:2015qya}
Z.~G.~He and B.~A.~Kniehl,
Complete Nonrelativistic-QCD Prediction for Prompt Double $J/\psi$ Hadroproduction,
Phys. Rev. Lett. \textbf{115}, no.2, 022002 (2015).
%doi:10.1103/PhysRevLett.115.022002
%[arXiv:1609.02786 [hep-ph]].
%27 citations counted in INSPIRE as of 29 Jul 2020


%\cite{Lansberg:2020rft}
\bibitem{Lansberg:2020rft}
J.~P.~Lansberg, H.~S.~Shao, N.~Yamanaka, Y.~J.~Zhang and C.~Noûs,
Complete NLO QCD study of single- and double-quarkonium hadroproduction in the colour-evaporation model at the Tevatron and the LHC,
Phys. Lett. B \textbf{807}, 135559 (2020).
%doi:10.1016/j.physletb.2020.135559
%[arXiv:2004.14345 [hep-ph]].
%0 citations counted in INSPIRE as of 02 Aug 2020


%\cite{Lansberg:2019fgm}
\bibitem{Lansberg:2019fgm}
J.~P.~Lansberg, H.~S.~Shao, N.~Yamanaka and Y.~J.~Zhang,
Prompt ${J/\psi}$-pair production at the LHC: impact of loop-induced contributions and of the colour-octet mechanism,
Eur. Phys. J. C \textbf{79}, no.12, 1006 (2019).
%doi:10.1140/epjc/s10052-019-7523-8
%[arXiv:1906.10049 [hep-ph]].
%8 citations counted in INSPIRE as of 02 Aug 2020


%\cite{Li:2009ug}
\bibitem{Li:2009ug}
R.~Li, Y.~J.~Zhang and K.~T.~Chao,
Pair Production of Heavy Quarkonium and $B_c^{(*)}$ Mesons at Hadron Colliders,
Phys. Rev. D \textbf{80}, 014020 (2009).
%doi:10.1103/PhysRevD.80.014020
%[arXiv:0903.2250 [hep-ph]].
%39 citations counted in INSPIRE as of 02 Aug 2020


%\cite{Aaij:2014bga}
\bibitem{Aaij:2014bga}
R.~Aaij \textit{et al.} [LHCb],
Measurement of the $\eta_c (1S)$ production cross-section in proton-proton collisions via the decay $\eta_c (1S) \rightarrow p \bar{p}$,
Eur. Phys. J. C \textbf{75}, no.7, 311 (2015).
%doi:10.1140/epjc/s10052-015-3502-x
%[arXiv:1409.3612 [hep-ex]].
%78 citations counted in INSPIRE as of 29 Jul 2020



%\cite{Aaij:2011sn}
\bibitem{Aaij:2011sn}
R.~Aaij \textit{et al.} [LHCb],
Observation of $X(3872) $ production in $pp$ collisions at $\sqrt{s}=7$ TeV,
Eur. Phys. J. C \textbf{72}, 1972 (2012).
%doi:10.1140/epjc/s10052-012-1972-7
%[arXiv:1112.5310 [hep-ex]].
%196 citations counted in INSPIRE as of 29 Jul 2020


%\cite{Bodwin:1994jh}
\bibitem{Bodwin:1994jh}
G.~T.~Bodwin, E.~Braaten and G.~P.~Lepage,
Rigorous QCD analysis of inclusive annihilation and production of heavy quarkonium,
Phys. Rev. D \textbf{51}, 1125-1171 (1995).
%doi:10.1103/PhysRevD.55.5853
%[arXiv:hep-ph/9407339 [hep-ph]].
%2433 citations counted in INSPIRE as of 29 Jul 2020


%\cite{Ma:2014mri}
\bibitem{Ma:2014mri}
Y.~Q.~Ma and R.~Venugopalan,
Comprehensive Description of $J/\psi$ Production in Proton-Proton Collisions at Collider Energies,
Phys. Rev. Lett. \textbf{113}, no.19, 192301 (2014).
%doi:10.1103/PhysRevLett.113.192301
%[arXiv:1408.4075 [hep-ph]].
%98 citations counted in INSPIRE as of 29 Jul 2020

%\cite{Li:2011yc}
\bibitem{Li:2011yc}
D.~Li, Y.~Q.~Ma and K.~T.~Chao,
$\chi_{cJ}$ production associated with a $c\bar c$ pair at hadron colliders,
Phys. Rev. D \textbf{83}, 114037 (2011).
%doi:10.1103/PhysRevD.83.114037
%[arXiv:1106.4262 [hep-ph]].
%13 citations counted in INSPIRE as of 29 Jul 2020



%\cite{Butenschoen:2014dra}
\bibitem{Butenschoen:2014dra}
M.~Butenschoen, Z.~G.~He and B.~A.~Kniehl,
$\eta_c$ production at the LHC challenges nonrelativistic-QCD factorization,
Phys. Rev. Lett. \textbf{114}, no.9, 092004 (2015).
%doi:10.1103/PhysRevLett.114.092004
%[arXiv:1411.5287 [hep-ph]].
%71 citations counted in INSPIRE as of 29 Jul 2020


%\cite{Han:2014jya}
\bibitem{Han:2014jya}
H.~Han, Y.~Q.~Ma, C.~Meng, H.~S.~Shao and K.~T.~Chao,
$\eta_c$ production at LHC and indications on the understanding of $J/\psi$ production,
Phys. Rev. Lett. \textbf{114}, no.9, 092005 (2015).
%doi:10.1103/PhysRevLett.114.092005
%[arXiv:1411.7350 [hep-ph]].
%61 citations counted in INSPIRE as of 29 Jul 2020


%\cite{Bodwin:2015iua}
\bibitem{Bodwin:2015iua}
G.~T.~Bodwin, K.~T.~Chao, H.~S.~Chung, U.~R.~Kim, J.~Lee and Y.~Q.~Ma,
Fragmentation contributions to hadroproduction of prompt$J/\psi$, $\chi_{cJ}$, and $\psi(2S)$ states,
Phys. Rev. D \textbf{93}, no.3, 034041 (2016).
%doi:10.1103/PhysRevD.93.034041
%[arXiv:1509.07904 [hep-ph]].
%51 citations counted in INSPIRE as of 29 Jul 2020


%\cite{Ma:2010vd}
\bibitem{Ma:2010vd}
Y.~Q.~Ma, K.~Wang and K.~T.~Chao,
QCD radiative corrections to $\chi_{cJ}$ production at hadron colliders,
Phys. Rev. D \textbf{83}, 111503 (2011).
%doi:10.1103/PhysRevD.83.111503
%[arXiv:1002.3987 [hep-ph]].
%134 citations counted in INSPIRE as of 29 Jul 2020


%\cite{Artoisenet:2009wk}
\bibitem{Artoisenet:2009wk}
P.~Artoisenet and E.~Braaten,
Production of the $X(3872)$ at the Tevatron and the LHC,
Phys. Rev. D \textbf{81}, 114018 (2010).
%doi:10.1103/PhysRevD.81.114018
%[arXiv:0911.2016 [hep-ph]].
%129 citations counted in INSPIRE as of 29 Jul 2020


%\cite{Butenschoen:2013pxa}
\bibitem{Butenschoen:2013pxa}
M.~Butenschoen, Z.~G.~He and B.~A.~Kniehl,
NLO NRQCD disfavors the interpretation of $X(3872)$ as $\chi_{c1}(2P)$,
Phys. Rev. D \textbf{88}, 011501 (2013).
%doi:10.1103/PhysRevD.88.011501
%[arXiv:1303.6524 [hep-ph]].
%34 citations counted in INSPIRE as of 29 Jul 2020



\iffalse

%\cite{Chao:1980dv}
\bibitem{Chao:1980dv}
K.~T.~Chao,
The (cc) - ($\bar{cc}$) (Diquark - Anti-Diquark) States in $e^+ e^-$ Annihilation,
Z. Phys. C \textbf{7}, 317 (1981).
%doi:10.1007/BF01431564
%29 citations counted in INSPIRE as of 27 Jul 2020



%\cite{Liu:2019zuc}
\bibitem{Liu:2019zuc}
M.~S.~Liu, Q.~F.~Lv, X.~H.~Zhong and Q.~Zhao,
All-heavy tetraquarks,
Phys. Rev. D \textbf{100}, no.1, 016006 (2019).
%doi:10.1103/PhysRevD.100.016006
%[arXiv:1901.02564 [hep-ph]].
%25 citations counted in INSPIRE as of 29 Jul 2020


%\cite{Chen:2016jxd}
\bibitem{Chen:2016jxd}
W.~Chen, H.~X.~Chen, X.~Liu, T.~G.~Steele and S.~L.~Zhu,
Hunting for exotic doubly hidden-charm/bottom tetraquark states,
Phys. Lett. B \textbf{773}, 247-251 (2017).
%doi:10.1016/j.physletb.2017.08.034
%[arXiv:1605.01647 [hep-ph]].
%49 citations counted in INSPIRE as of 27 Jul 2020


%\cite{Wang:2019rdo}
\bibitem{Wang:2019rdo}
G.~J.~Wang, L.~Meng and S.~L.~Zhu,
Spectrum of the fully-heavy tetraquark state $QQ\bar Q' \bar Q'$,
Phys. Rev. D \textbf{100}, no.9, 096013 (2019).
%doi:10.1103/PhysRevD.100.096013
%[arXiv:1907.05177 [hep-ph]].
%16 citations counted in INSPIRE as of 27 Jul 2020


%\cite{Wang:2021kfv}
\bibitem{Wang:2021kfv}
G.~J.~Wang, L.~Meng, M.~Oka and S.~L.~Zhu,
Higher fully charmed tetraquarks: Radial excitations and P-wave states,
Phys. Rev. D \textbf{104}, no.3, 036016 (2021).
%doi:10.1103/PhysRevD.104.036016
%[arXiv:2105.13109 [hep-ph]].
%11 citations counted in INSPIRE as of 10 Jul 2022


\fi


%\cite{Dong:2020hxe}
\bibitem{Dong:2020hxe}
X.~K.~Dong, F.~K.~Guo and B.~S.~Zou,
Explaining the Many Threshold Structures in the Heavy-Quark Hadron Spectrum,
Phys. Rev. Lett. \textbf{126}, no.15, 152001 (2021).
%doi:10.1103/PhysRevLett.126.152001
%[arXiv:2011.14517 [hep-ph]].
%62 citations counted in INSPIRE as of 31 Aug 2022


%\cite{Lyu:2021qsh}
\bibitem{Lyu:2021qsh}
Y.~Lyu, H.~Tong, T.~Sugiura, S.~Aoki, T.~Doi, T.~Hatsuda, J.~Meng and T.~Miyamoto,
Dibaryon with Highest Charm Number near Unitarity from Lattice QCD,
Phys. Rev. Lett. \textbf{127}, no.7, 072003 (2021).
%doi:10.1103/PhysRevLett.127.072003
%[arXiv:2102.00181 [hep-lat]].
%17 citations counted in INSPIRE as of 10 Jul 2022



\iffalse


%\cite{Berezhnoy:2011xn}
\bibitem{Berezhnoy:2011xn}
A.~V.~Berezhnoy, A.~V.~Luchinsky and A.~A.~Novoselov,
Tetraquarks Composed of 4 Heavy Quarks,
Phys. Rev. D \textbf{86}, 034004 (2012).
%doi:10.1103/PhysRevD.86.034004
%[arXiv:1111.1867 [hep-ph]].
%48 citations counted in INSPIRE as of 29 Jul 2020


%\cite{Karliner:2016zzc}
\bibitem{Karliner:2016zzc}
M.~Karliner, S.~Nussinov and J.~L.~Rosner,
$Q Q \bar Q \bar Q$ states: masses, production, and decays,
Phys. Rev. D \textbf{95}, no.3, 034011 (2017).
%doi:10.1103/PhysRevD.95.034011
%[arXiv:1611.00348 [hep-ph]].
%66 citations counted in INSPIRE as of 27 Jul 2020


%\cite{Karliner:2017qhf}
\bibitem{Karliner:2017qhf}
M.~Karliner, J.~L.~Rosner and T.~Skwarnicki,
Multiquark States,
Ann. Rev. Nucl. Part. Sci. \textbf{68}, 17-44 (2018).
%doi:10.1146/annurev-nucl-101917-020902
%[arXiv:1711.10626 [hep-ph]].
%87 citations counted in INSPIRE as of 29 Jul 2020


%\cite{Berezhnoy:2011xy}
\bibitem{Berezhnoy:2011xy}
A.~V.~Berezhnoy, A.~K.~Likhoded, A.~V.~Luchinsky and A.~A.~Novoselov,
Double $J/\psi$-meson Production at LHC and 4c-tetraquark state,
Phys. Rev. D \textbf{84}, 094023 (2011).
%doi:10.1103/PhysRevD.84.094023
%[arXiv:1101.5881 [hep-ph]].
%101 citations counted in INSPIRE as of 29 Jul 2020


%\cite{Richard:2017vry}
\bibitem{Richard:2017vry}
J.~M.~Richard, A.~Valcarce and J.~Vijande,
String dynamics and metastability of all-heavy tetraquarks,
Phys. Rev. D \textbf{95}, no.5, 054019 (2017).
%doi:10.1103/PhysRevD.95.054019
%[arXiv:1703.00783 [hep-ph]].
%47 citations counted in INSPIRE as of 29 Jul 2020



%\cite{Czarnecki:2017vco}
\bibitem{Czarnecki:2017vco}
A.~Czarnecki, B.~Leng and M.~B.~Voloshin,
Stability of tetrons,
Phys. Lett. B \textbf{778}, 233-238 (2018).
%doi:10.1016/j.physletb.2018.01.034
%[arXiv:1708.04594 [hep-ph]].
%43 citations counted in INSPIRE as of 29 Jul 2020

\fi






\end{thebibliography}
\end{document}